\documentclass[a4paper,11pt]{article}

\usepackage{amsfonts,amssymb,amsmath,latexsym}
\usepackage{graphicx,graphics}
\usepackage{epsfig}
\usepackage{psfrag}
\usepackage{tabls,caption}
\usepackage{rotating}
\usepackage{mathrsfs}
\usepackage{booktabs}
\usepackage{threeparttable}
\usepackage{chapterbib}
\usepackage{color}
\usepackage[semicolon]{natbib}
\usepackage[gen]{eurosym}
\usepackage{hyperref}
\usepackage{comment}
\usepackage{ragged2e}

\newcommand{\rN}{\mathcal{N}}

\begin{document}

\title{SHARELIFE Imputations\thanks{
\ Corresponding author: giuseppe.deluca@unipa.it.
We acknowledge financial support from
`Challenge 4 - Trajectories for Active and Healthy Ageing (Behavioural and Psychological Determinants)',
under the Research and Innovation Programme of the Extended Partnership `AGE--IT', funded through the resources of the National Recovery and Resilience Plan (PNRR), Mission 4 `Education and Research', Component 2 `From Research to Business', Investment Line 1.3 `Extended Partnerships among Universities, Research Centres and Companies for the Funding of Basic Research Projects' Call
(code PE00000015, CUP E63C22002050006), Theme 1.5, Project title: SHARELIFE-MI (n.\ 1673755).
We are grateful to Guglielmo Weber helpful discussions.}}
\author{
Giuseppe De Luca\\
\emph{\small{University of Palermo, Italy}}\\
\vspace{5mm}\\
Paolo Li Donni\\
\emph{\small{University of Palermo, Italy}}\\
\vspace{5mm}\\
}
\date{\today}

\maketitle

\vfill

\begin{abstract}
This report describes the SHARELIFE-MI project,
which aims to generate multiple imputations for missing values in the life-course data collected in SHARELIFE Waves 3 and 7.
The SHARELIFE study reconstructs individual life histories through retrospective questions covering key biographical domains
such as partnerships, fertility, employment, and residence.
As in the regular SHARE waves, item nonresponse represents an important source of nonsampling error---particularly for
monetary variables, which require conversions across multiple currencies and long time periods.
We document the preliminary data recoding and harmonization steps,
as well as the design, specification, and implementation of an imputation model based on the fully conditional specification approach.
Finally, we assess the internal and external validity of the resulting imputations through comparisons with the observed data,
alternative nonresponse adjustments based on inverse propensity weighting, and external benchmarks from the regular SHARE waves.
\end{abstract}

\vfill

\begin{center}

{\bf Keywords}: SHARELIFE, item nonresponse, mutiple imputations,
fully conditional specification,
internal and external validity

\bigskip


\end{center}

\vfill

\thispagestyle{empty}

\setcounter{page}{1}
\baselineskip 19pt

\newpage
\section{Introduction}

Ideally, analyzing the long-term effects of social policy reforms would require longitudinal data that follow individuals from school-leaving age through old age. However, collecting such data prospectively is often infeasible due to time and cost constraints. As an alternative, several surveys of the elderly population gather life-history data retrospectively by asking respondents to recall and describe their living conditions at various points in their lives.
Examples include the Health and Retirement Study (HRS) in the United States, the English Longitudinal Study of Ageing (ELSA) in the United Kingdom, and the Survey of Health, Ageing and Retirement in Europe (SHARE).
The common goal of the life-history components in these studies is to reconstruct respondents’ life courses prior to the baseline interview of each survey.
In what follows, we focus on the third and seventh waves of SHARE---collectively known as SHARELIFE---which provide rich retrospective information on childhood conditions and detailed life-course histories of family, housing, health, and employment for individuals aged 50 and older across 27 European countries and Israel.

Retrospective data on early life events are of great value to social scientists,
as they help overcome several key limitations of standard panel data.
Nevertheless, skepticism persists regarding the ability of older respondents to
accurately recall events that occurred many years earlier.
Despite encouraging evidence from previous validation studies (see, e.g., \citealt{Havari_Mazzonna_2015}),
major concerns remain for questions involving monetary variables---such as maternity benefits,
wages and self-employment income at the beginning of job spells,
pension benefits upon retirement,
current wages and self-employment income, and earnings at the end of the main job.
For each monetary item,
SHARELIFE respondents are asked to provide information on the reference period of a given spell,
the corresponding monetary amount, and the currency in which the amount was received.
Monetary data are considered valid only if all three questions are answered.
Moreover, measurement errors in any of these components (time period, amount, or currency)
can easily generate outliers that exert a disproportionate influence on survey estimates.

One implication of this retrospective data collection process is that the monetary variables
collected in SHARELIFE exhibit particularly low item response rates, ranging from a minimum of about 34 percent for
maternity benefits to a maximum of around 82 percent for current monthly wages.
In addition, the harmonization of these monetary amounts requires complicated
conversion strategies across multiple currencies and long time periods.

Our project aims to enhance the usability of these retrospective data by generating multiple imputations
for the missing values in SHARELIFE Waves 3 and 7.
The resulting SHARELIFE-MI database contains five multiple imputations for approximately 650 variables.
These variables cover all key domains of the SHARELIFE interview, including socio-demographic characteristics,
family and partnership histories, health, employment, accommodation, and monetary outcomes.
To maximize comparability, all monetary amounts are expressed in 2017 euros and adjusted for purchasing power
parity (PPP).

Multiple imputations are designed to address selection bias and the loss of precision caused by item nonresponse
and outliers in the PPP-adjusted distribution of monetary variables, under certain assumptions about the missing-data
mechanism and the model used to generate imputed values (the imputation model).
Following \citet{Rubin_1976}, three types of missing-data mechanisms are typically distinguished:
missing completely at random (MCAR), missing at random (MAR), and missing not at random (MNAR).
A mechanism is MCAR if the probability of missingness is independent of both observed and unobserved variables;
MAR if it depends only on observed variables;
and MNAR if it depends on unobserved variables even after conditioning on the observed data.
Standard applications of multiple imputations assume that missingness is MAR and that the imputation model is
congenial with the analysis model, in the sense of \citet{Meng_1994}.
These assumptions can be relaxed using more advanced approaches for analysis of imputed data, such as the
generalized missing indicator approach of
\citet[\citeyear{Dardanoni_EtAl_2015}]{Dardanoni_EtAl_2011}.

The remainder of the report is organized as follows.
Section~\ref{sec:SL_data} presents an overview of the SHARELIFE data.
Section~\ref{sec:money_conv} describes the monetary conversion strategies employed to harmonize amounts across countries and over time.
Section~\ref{sec:descriptive} discusses descriptive statistics on response rates and missing-data patterns.
Section~\ref{sec:MI} provides a brief introduction to multiple imputation methodology, with a special focus on the fully conditional specification method of \citet{VanBuuren_EtAl_1999}, relevant generalizations of this approach, and the specification and implementation of the SHARELIFE imputation model.
Section~\ref{sec:validation} presents diagnostic analyses of the internal and external validity of the multiple imputations for monetary variables.
Section~\ref{sec:CONCLUSIONS} concludes.
The report also includes three appendices.
Appendix~A provides detailed information on the specification of the SHARELIFE imputation models;
Appendix~B presents scatterplots of the monetary variables; and
Appendix~C offers additional analyses of the determinants of the missing-data processes and nonresponse adjustments based on the propensity score.


\section{SHARELIFE data}
\label{sec:SL_data}

Our project uses data from SHARE release 9.0.0, a multidisciplinary and cross-national panel survey that covers nationally representative samples of the elderly population aged 50 and older in 27 European countries and Israel.
In addition to specific studies such as the two SHARE Corona surveys (\citealt{Bergmann_Borsch-Supan_2021}) and the SHARE Harmonised Cognitive Assessment Protocol (SHARE-HCAP; \citealt{Bergmann_EtAl_2024}; \citealt{Douhou_EtAl_2025}), the SHARE panel currently consists of seven regular waves (2004, 2006, 2011, 2013, 2015, 2019, 2021) and two retrospective waves (2008, 2017).
The regular waves collect longitudinal data on respondents’ lives at approximately 2-year intervals, measuring physical and mental health, economic and social activities, income and wealth, consumption and healthcare expenditures, behavioral risks, expectations, transfers of time and money within and outside the family, as well as levels of satisfaction and well-being.
The retrospective waves, known as SHARELIFE, instead collect data on respondents’ entire life course, from childhood up to the year of the interview, covering fertility and children’s characteristics, cohabitation and housing choices, education, socioeconomic and occupational conditions, job characteristics, income, disability benefits, health status and healthcare, and other general living conditions.

Sampling design procedures are managed separately by country, but in accordance with common standards aimed at preserving the representativeness of national samples.
The SHARE panel does not rely on rotation methods and is periodically updated with refreshment samples to ensure coverage of younger cohorts while at the same time mitigating the effects of attrition on sample size.
All interviews are conducted through CAPI (computer-assisted personal interview) using standardized multilingual questionnaires across countries.

Like all voluntary sample surveys, each wave of the SHARE panel is subject to both sampling and non-sampling errors.
To address these issues, the survey provides design weights to correct for unequal household inclusion probabilities, calibrated weights to adjust for unit nonresponse and attrition, and multiple imputations to handle item nonresponse.
Weights are available for all waves, whereas imputations are only provided for data collected in the regular waves.
The SHARELIFE-MI project aims to generate multiple imputations for the missing values of the SHARELIFE interviews of waves 3 and 7.

\begin{table}[htp]
\caption{SHARELIFE sample by country and wave}
\begin{center}
\begin{tabular}{crrr}
\hline
\multicolumn{1}{l}{Country} &
\multicolumn{1}{c}{Wave 3}  &
\multicolumn{1}{c}{Wave 7}  &
\multicolumn{1}{c}{Total} \\
\hline
     AT     &     993     &   2,693     &   3,686    \\
     BE     &   2,865     &   3,333     &   6,198    \\
     BG     &       0     &   1,998     &   1,998    \\
     CH     &   1,324     &   1,648     &   2,972    \\
     CY     &       0     &   1,233     &   1,233    \\
     CZ     &   1,809     &   3,289     &   5,098    \\
     DE     &   1,918     &   2,982     &   4,900    \\
     DK     &   2,143     &   1,961     &   4,104    \\
     EE     &       0     &   5,115     &   5,115    \\
     ES     &   2,271     &   3,423     &   5,694    \\
     FI     &       0     &   2,007     &   2,007    \\
     FR     &   2,500     &   2,186     &   4,686    \\
     GR     &   3,090     &   1,160     &   4,250    \\
     HR     &       0     &   2,408     &   2,408    \\
     HU     &       0     &   1,538     &   1,538    \\
     IE     &     855     &       0     &     855    \\
     IL     &       0     &   2,131     &   2,131    \\
     IT     &   2,528     &   2,997     &   5,525    \\
     LT     &       0     &   2,035     &   2,035    \\
     LU     &       0     &   1,250     &   1,250    \\
     LV     &       0     &   1,734     &   1,734    \\
     MT     &       0     &   1,261     &   1,261    \\
     NL     &   2,258     &       0     &   2,258    \\
     PL     &   1,939     &   3,553     &   5,492    \\
     PT     &       0     &   1,282     &   1,282    \\
     RO     &       0     &   2,114     &   2,114    \\
     SE     &   1,961     &   2,127     &   4,088    \\
     SI     &       0     &   3,691     &   3,691    \\
     SK     &       0     &   2,077     &   2,077    \\
\hline
    Total   &  28,454     &  63,226     &  91,680    \\
\hline
\end{tabular}
\end{center}
\label{tab:sample}
\end{table}

Table~\ref{tab:sample} reports the SHARELIFE sample by country and wave, amounting to 91,680 retrospective interviews across 28 countries.
Wave 3 (2008–09) introduced the first SHARELIFE interview, covering 28,454 respondents from 14 countries
(Austria, Belgium, Czech Republic, Denmark, France, Germany, Greece, Ireland, Italy, the Netherlands, Poland, Spain, Sweden, and Switzerland)
who had already participated in one of the first two regular waves.
The subsequent three regular waves (2011, 2013, and 2015) extended the study to additional countries
(Bulgaria, Cyprus, Croatia, Estonia, Finland, Hungary, Israel, Latvia, Lithuania, Luxembourg, Malta, Romania, Slovak Republic, and Slovenia) and incorporated refreshment samples within existing ones.
In Wave 7 (2017), SHARELIFE interviews were administered to 63,226 respondents who had not participated in Wave 3, while those who had already completed a SHARELIFE interview in Wave 3 received the standard questionnaire.

\begin{table}[htp]
\caption{Flow of SHARELIFE interview modules}
\begin{center}
\begin{tabular}{lll}
\hline
\multicolumn{2}{c}{Module}  &
\multicolumn{1}{l}{ }  \\
\multicolumn{1}{l}{Wave 3}  &
\multicolumn{1}{c}{Wave 7}  &
\multicolumn{1}{c}{Description}  \\
\hline
CV\_R & CV\_R & Coverscreen \\
ST  & DN & Demographics\\
RC & RC & Retrospective children  \\
RP & RP & Retrospective partner \\
AC & RA & Retrospective accommodation\\
CS & CC & Childhood circumstances \\
RE & RE & Retrospective employment \\
WQ & WQ & Work quality \\
DQ & DQ & Disability \\
FS & FS & Financial section \\
HS & HS & Health section \\
HC & RH & Retrospective health care \\
GL & GL & General life and Persecution \\
GS & GS & Grip strength \\
IV & IV & Interviewer observations	\\
\hline
\end{tabular}
\end{center}
\label{tab:int-modules}
\end{table}

Table~\ref{tab:int-modules} presents the flow of the SHARELIFE interview modules.
As in the regular SHARE waves, the interview begins with a coverscreen (CV\_R) module on family composition, followed by a demographic module (ST/DN) that collects baseline socio-demographic information.
The retrospective part of the SHARELIFE interview starts with the children (RC) module,
which records information on natural and adopted children, including births, characteristics, maternity leave, and maternity benefits.
This is followed by the partner (RP) module on partnership histories ---covering cohabitation, marriages, and divorces --- and the accommodation (AC/RA) module on past residences, moves, housing types, and ownership.
The childhood circumstances (CS/CC) module then collects information on living conditions around age 10, including housing, health, school performance, and socio-economic background.
The employment (RE) module documents detailed work histories, job characteristics, incomes, retirement decisions, and pension benefits, with complementary questions from the work quality (WQ) and disability (DQ) modules on job demands, control, work effort, disability leaves, work reductions, and disability pensions.
Subsequent modules address finances, health, and broader life experiences. %
The financial section (FS) covers past investments, such as stocks, funds, insurance, and retirement savings.
The health section (HS) records major life-course health events, including hospitalizations, illnesses, and diseases.
The health care (HC/RH) module asks about services received ---vaccinations, doctor visits, preventive check-ups, and health behaviors.
The general life and persecution (GL) module collects information on periods of happiness, stress, hardship, and hunger, as well as experiences of discrimination, persecution, and oppression by respondents or their parents.
The interview concludes with a physical grip strength assessment and interviewer observations on the interview process.

As discussed in \citet{Schroder_2011},
the interview mode used in SHARELIFE combines computer-assisted personal interviewing (CAPI) with event history calendars (EHC).
CAPI is widely recognized as one of the most effective interviewing methods, particularly for reducing item nonresponse and measurement errors.
The EHC approach, also known as the life calendar method, enhances the collection of retrospective data by drawing on insights from cognitive psychology
and memory research.
Specifically, life events are recorded within a large grid that combines the time dimension along the horizontal axis with key life domains---such as children, partnerships, and employment—along the vertical axis.
This structure enables respondents to visualize important events from different areas of their lives in parallel, thereby facilitating recall and improving data consistency.
Unlike the standard SHARE interview, however, the SHARELIFE questionnaire does not include sequences of unfolding bracket questions for nonrespondents to open-ended monetary items.
The absence of such partial information introduces additional uncertainty into the imputation of missing monetary amounts.

\section{Money conversion strategies}
\label{sec:money_conv}

\subsection{Structure of monetary variables}\label{sec:money_var}
Imputing monetary variables is arguably one of the most complex tasks when working with SHARELIFE data.  These variables record self-reported amounts across multiple domains on respondents' job histories, pension and maternity benefits. Monetary amounts span about 148 distinct currencies. In our data, the earliest observations date to the early 1920s (e.g., French francs) and the latest to 2019.

In the SHARELIFE data we focus on the following self-reported amounts: maternity benefits for adopted and natural children (available only for women), first monthly wage and income, current monthly wage and income, and finally the wage and income at the end of the first job. For each of the self-reported amounts, respondents provide the monetary amount received, the currency, and the year of receipt. In the case of maternity benefits, the reference year is anchored to the child’s birth year, which is treated as the first year in which the benefit is received.
Notice that first monthly wages/income from work are spell data, while pension benefits when retired have been computed by pooling all pension benefits received from each job spell. In general, monetary items are collected as follows: respondents first state the amount, then the reference year, and finally the currency in which the amount has been reported.

SHARELIFE records three kinds of currency labels: specific ISO or legacy codes; generic labels that compress several historical currencies under a single name; and explicit historical series that refer to well-defined institutional periods (e.g., Soviet rubles, Czechoslovak koruna, Israeli pound). 

\begin{table}[htbp]
\caption{Group 1 currencies: counts by monetary item}
\label{tab:curr-bytype-gr1}
\begin{center}
\begin{tabular}{lrrrr}
\hline
\multicolumn{1}{c}{Currency} &
\multicolumn{1}{c}{$Y_1$} &
\multicolumn{1}{c}{$Y_2$} &
\multicolumn{1}{c}{$Y_3$} &
\multicolumn{1}{c}{$Y_4$} \\
\hline
EUR -- Euro                   & 148 & 6,116 & 11,441 & 9,336 \\
ATS -- Austrian schilling     & 365 & 2,289 &     3 &  859 \\
BEF -- Belgian franc          & 207 & 3,976 &    43 & 1,253 \\
DEM -- German mark (FRG)      & 269 & 4,100 &    14 & 1,358 \\
DKK -- Danish krone           & 123 & 3,017 &  1,593 & 1,112 \\
ESP -- Spanish peseta         &  16 & 2,278 &   100 &  852 \\
FIM -- Finnish markka         & 106 & 1,208 &    11 &  161 \\
FRF -- French franc           &  89 & 2,549 &    21 &  702 \\
ITL -- Italian lira           &  98 & 3,372 &    10 & 1,384 \\
LUF -- Luxembourg franc       &  43 &  755 &     5 &  195 \\
NLG -- Dutch guilder          &   2 &  124 &     0 &   35 \\
PTE -- Portuguese escudo      &  15 &  729 &     2 &  191 \\
IEP -- Irish pound            &   4 &  292 &     2 &   74 \\
GRD -- Greek drachma          &  26 & 1,211 &     5 &  405 \\
CYP -- Cyprus pound           &  29 &  809 &    12 &  274 \\
MTL -- Maltese lira           &   2 &  831 &    10 &  355 \\
CHF -- Swiss franc            &  43 & 2,733 &   738 & 1,415 \\
GBP -- Pound sterling         &   6 &  276 &     1 &   38 \\
\hline
\end{tabular}
\vspace{-0.7cm}
\parbox{102mm}{\footnotesize
Note: $Y_1$ = Monthly maternity benefits;
$Y_2$ = First monthly wage; $Y_3$ = Current monthly wage;
$Y_4$ = Wage at the end of the main job.}
\end{center}
\end{table}

\begin{table}[htp]
\caption{Group 2 (CEE transitions and redenominations): counts by monetary item}
\begin{center}
\begin{tabular}{lrrrrr}
\hline
\multicolumn{1}{c}{Currency} &
\multicolumn{1}{c}{$Y_1$} &
\multicolumn{1}{c}{$Y_2$} &
\multicolumn{1}{c}{$Y_3$} &
\multicolumn{1}{c}{$Y_4$} \\
\hline
BGL -- Bulgarian leva                               & 579 & 1728 &  512 & 1190 \\
CSK -- Czechoslovak koruna                          & 265 &  773 &   30 &  297 \\
CZK -- Czech koruna                                 & 1188 & 3799 &  760 & 2755 \\
DDM -- East German mark (GDR)                       &  88 &  841 &    1 &  131 \\
EEK -- Estonian kroon                               &  28 & 1208 &    8 &  650 \\
HRD -- Croatian dinar                               &  18 &   82 &    0 &   56 \\
HRK -- Croatian kuna                                &  84 &  513 &  354 &  810 \\
HUF -- Hungarian forint                             & 257 & 1266 &  162 &  907 \\
LTL -- Lithuanian litas                             &  47 &  562 &    7 &  441 \\
LTT -- Lithuanian talonas                           &  12 &   61 &    0 &   33 \\
LVL -- Latvian lats                                 &  18 &  481 &    6 &  335 \\
LVR -- Latvia ruble                                 &   1 &   95 &    2 &   36 \\
PLN -- Polish zloty (new)                           &  38 &  945 &  777 &  828 \\
PLZ -- Polish zloty (old) — explicit                &  98 & 1508 &    9 &  279 \\
ROL -- Romanian leu (old)                           &  63 & 1391 &   18 &  797 \\
RON -- Romanian leu (new)                           &   7 &  248 &  281 &  296 \\
SIT -- Slovenian tolar                              &   7 &  131 &    1 &  256 \\
SKK -- Slovak koruna                                & 126 &  424 &    5 &  350 \\
SUR -- Soviet rouble                                &   0 &    4 &    0 &    0 \\
RUB -- Russian ruble                                &   7 &   83 &    0 &   23 \\
Soviet rubles (1940–1941) -- explicit               &  71 & 1353 &    0 &  315 \\
Soviet rubles (1944–1992) -- explicit               &   2 &   33 &    0 &    3 \\
YUD -- Yugoslav new dinar                           & 109 &  793 &    0 &  138 \\
YUN -- Yugoslav dinar                               &   0 &    2 &    0 &    1 \\
Czechoslovak koruna (1919–1939) -- explicit         &   1 &    2 &    0 &    0 \\
Czechoslovak koruna (1945–1953) -- explicit         &   5 &   39 &    0 &    2 \\
Czechoslovak koruna (1953–1992) -- explicit         & 639 & 1877 &    4 &  164 \\
Estonian old kroon (1928–1940) -- explicit          &   0 &   13 &    0 &    2 \\
Estonian mark -- explicit                           &   0 &    1 &    0 &    0 \\
Hungarian peng\H{o} -- explicit                         &   0 &    6 &    0 &    2 \\
Slovenian provisional payment notes -- explicit     &   0 &    4 &    0 &    7 \\
First leu (1867–1944) -- explicit                   &   1 &   67 &    0 &   26 \\
Second leu (1947–1952) -- explicit                  &   0 &    5 &    0 &    1 \\
Red Army occupation leu (1944–1947) -- explicit     &   0 &    2 &    0 &    0 \\
PLN -- ``explicitly PLN: new Polish zloty''           &   6 &   68 &   20 &   46 \\
\hline
\end{tabular}
\vspace{-0.7cm}
\parbox{133mm}{\footnotesize
Note: $Y_1$ = Monthly maternity benefits; $Y_2$ = First monthly wage; $Y_3$ = Current monthly wage;
$Y_4$ = Wage at the end of the main job.}
\end{center}
\label{tab:curr-bytype-gr2}
\end{table}

\begin{table}[htp]
\caption{Group 3 (extra-EU currencies): counts by monetary item}
\begin{center}
\begin{tabular}{p{8cm}rrrr}
\hline
\multicolumn{1}{c}{Currency} &
\multicolumn{1}{c}{$Y_1$}  &
\multicolumn{1}{c}{$Y_2$} &
\multicolumn{1}{c}{$Y_3$} &
\multicolumn{1}{c}{$Y_4$} \\
\hline
CAD -- Canadian dollar        &   1 &  43 &   0 &   7 \\
ILS -- Israeli new shekel     &  12 & 570 & 398 & 490 \\
USD -- United States dollar   &   0 & 130 &   1 &  27 \\
ZAR -- South African rand     &   1 &  14 &   0 &   3 \\
\hline
\multicolumn{5}{l}{\emph{Explicit Israeli historical series}} \\
Israeli pound (lira), 1952--1980      &   3 & 319 &   0 &  11 \\
Palestine pound, 1948--1951           &   0 &  25 &   0 &   0 \\
Israeli old shekel, 1980--1985        &   3 &  23 &   0 &   3 \\
\hline
\end{tabular}
\vspace{-0.7cm}
\parbox{120mm}{\footnotesize
Note: $Y_1$ = Monthly maternity benefits; $Y_2$ = First monthly wage; $Y_3$ = Current monthly wage; $Y_4$ = Wage at the end of the main job.}
\end{center}
\label{tab:curr-bytype-gr3}
\end{table}

Viewed through a historical lens, 149 distinct currencies can be naturally grouped into three groups: (i) euro and pre-euro legacy currencies; (ii) transition and redenominated currencies in Central and Eastern Europe (CEE); and (iii) extra-EU currencies linked to migration or overseas spells. Finally Generic and explicit historical labels (e.g., [generic] rubel, krona/kroner/kronen, Reichsmark) cut across these groups and must be normalised to canonical codes using the spell's country and year of the interview. The currency in the three groups are presented in Table \ref{tab:curr-bytype-gr1}, \ref{tab:curr-bytype-gr2} and \ref{tab:curr-bytype-gr3}. Currencies in Group 3 (extra-EU) presented in Table \ref{tab:curr-bytype-gr3} were trimmed to codes with at least ten observations across the four monetary items, while all Israeli historical series were retained in full due to their clear internal timeline: first the Palestine pound, then the Israeli pound, followed by the old shekel, and finally the new shekel (ILS).

Reading across Group 1 (Table \ref{tab:curr-bytype-gr1}), a clear life-cycle pattern emerges. Current monthly wages are concentrated in currencies that either survived into the late stages of respondents’ careers or transitioned smoothly into the euro; EUR dominates this column. Among legacy issuers, DKK and SEK stand out with many current-wage observations, consistent with non-adoption of the euro and long, continuous series. First monthly wages, by contrast, lean heavily toward pre-euro stalwarts—ATS, BEF, FRF, ITL, DEM—which is exactly what we expect for earlier job starts. The Czech koruna (CZK) shows strong counts in both first and current wages, a signature of long working lives spanning the 1990s–2000s. Smaller but telling volumes in IEP, GRD, CYP, MTL reflect late-adopting or small economies where early-career spells precede euro cash introduction. Maternity benefits trace a different geography. They are present across Western legacies but are less dominant than for first wages, hinting at cohort timing and administrative practices that made benefits more salient in other regions. In sum, Group 1 validates the expected transition: early careers in national currencies, late careers increasingly in EUR, with robust current-wage mass where currency regimes remained stable (DKK, SEK) or migrated cleanly to the euro.

In Group 2 (Table \ref{tab:curr-bytype-gr2}), the currency counts read like a timeline of institutional change. CZK is prominent in both first and current wages, reflecting careers that bridge the Czechoslovak dissolution and the 1990s expansion. PLZ vs PLN clearly marks the 1995 redenomination: first wages are more often in PLZ, while current wages lean toward PLN. ROL vs RON shows the same logic for Romania’s 2005 reform, with older spells in ROL and more recent ones in RON. The Baltic trio—EEK, LTL/LTT, LVL/LVR—maps the post-Soviet re-establishment of national currencies and the later move to the euro. DDM appears alongside DEM, reminding us of the two Germanys prior to unification. Smaller but meaningful volumes in HUF, HRK, SKK and the explicit Soviet ruble series anchor the region’s transition from centrally planned regimes to market currencies. Maternity benefits in Group 2 are sizeable for CZK, HUF, BGL, pointing to earlier family-formation timing and institutional channels that made these transfers visible and reportable \citep{Sobotka_2004}. Together, the counts underscore why canonical mapping and redenomination handling are not cosmetic: they are essential to align each record to the correct year and currency before proceeding with any conversion.

Finally, the generic labels deserves a brief comment. Among entries tagged as ``[generic]'', the ``[generic] rubel'' carries thousands of first-wage entries; similar, though smaller, effects appear for ``[generic] gulden'', ``[generic] pounds'', and the krona/kroner/kronen variants. These are not misreported currency-they compress multiple historical currencies under one word. Without a country-by-year crosswalk, any attempt to retrieve the appropriate annual exchange rate would be unreliable.

\subsection{Money conversion across countries and over time}
\label{sec:money_cov}

As described in the previous section, the dataset spans many countries and currencies over a long time period, encompassing structural changes in currency regimes and institutional arrangements driven by wars and political transitions. These features pose several challenges for imputation. The first challenge is imputing missing amounts across heterogeneous currencies and countries. Once the data are partitioned into relevant cells (country/year/ monetary item/currency), the number of observations available for reliable imputation can be very limited. In such sparse settings, standard imputation strategies become fragile and may border on arbitrariness, as they are driven more by sampling noise than by signal.

Second, currency labels are heterogeneous. When amounts are reported with precise ISO or legacy codes, the link between amount, year, and currency is straightforward; with generic or explicit historical labels, that link is not one-to-one unless we also condition on the respondent’s country and the reference year. This is why every record must be resolved to a canonical currency-year before any modelling. For example entries labelled as ``[generic] francs'' do not identify a single currency. In SHARELIFE this tag may correspond to the French franc (FRF), the Belgian franc (BEF), or the Luxembourg franc (LUF). Similarly, the label ``[generic] dinar'' compresses multiple historical currencies. In the former Yugoslav area, pre-breakup spells map to the Yugoslav dinar (YUD/YUN) by year, while successor states adopt their post-1990s currencies (e.g., HRD/HRK for Croatia, SIT for Slovenia, BAM for Bosnia and Herzegovina). 

Third, redenominations and regime breaks create discontinuities in nominal series that can masquerade as economic jumps. Particularly, a redenomination is an official change in the face value of a currency that removes zeros from banknotes and accounting units without altering purchasing power. Statistically, it means that the numerical scale changes at a known date by a known factor; the underlying real value does not. Fo example, on 1 January 1995, Poland replaced the old złoty (PLZ) with the new złoty (PLN) at the official factor: 10,000 PLZ = 1 PLN. Taking into account this issue is crucial to prevent artificial jumps and to keep the self-reported amounts' series on a consistent monetary scale.

Fourth, gaps due to instability in exchange-rate coverage-especially during wars and high-inflation episodes—mean that a ``direct'' conversion is not always available or reliable. 
Fifth, conceptual mismatches and recall. Monetary variables differ in timing and definition (first wage vs current wage; wage vs income; gross vs net). Respondents may round or heap values, and recall error grows with the age of the spell. If we ignore these differences, imputation models will over-borrow information across unlike quantities. 

Finally, migration and foreign-currency earnings. A non-trivial share of records reflects work abroad (USD, CAD, ILS, etc.). Without a proper country-year crosswalk and conversion backbone, these observations can distort country-specific donor pools. 
Sixth, distributional tails and outliers. Hyperinflation years, miskeyed zeros, and rare currencies inflate the upper tail. Imputation methods are sensitive to such extremes. 

\subsection{Our approach to this problem}
\label{sec:money_approach}

The issues described above have clear implications for the imputation design. First, in a historically heterogeneous, multi-country and multi-currency setting, it is preferable to express all monetary information on a single reference scale (one currency and one base year) before imputing amounts. This standardization avoids cascading imputations -i.e., having to impute the reference year or the currency first - and places all records on a comparable footing. Standardizing all records to a single currency and base year before imputing does not create ``new'' information - amounts are merely re-expressed - but it increases the effective information available to the prediction model by enabling the use of partially observed cases and borrowing strength across comparable cells.

Second, each observation must be mapped to a canonical currency–year to avoid mixing non-comparable nominal series. Finally, all amounts must be converted to a common scale and, where cross-country comparability is required, expressed in PPP terms. Third, we need a transparent strategy for gaps in conversion coverage and for instability due to wars or historical financial crises—either by using a CPI-assisted route, a PPP-only path, or documented reconstructions of fixed-parity regimes.

The strategy proceeds in three steps. First, for each respondent we map the self-reported amount to a specific currency label. If the label recorded in the interview is a precise ISO or legacy code and is consistent with the reported year, we keep it as is. If the label is not specific, we determine the appropriate currency by jointly using consistently the respondent's country and the reference year. For example, if an Austrian interview reports a maternity-benefit amount as ``shilling'' in year $t$, we code it as Austrian schilling for the corresponding year $t$. This requires a currency–year crosswalk that checks interval consistency: the assigned currency must be the one legally in force in that country in year $t$.

In the second step we convert this one-to-one currency–year mapping into a common unit—US dollars—using historical exchange rates; we then re-express all amounts in constant 2017 USD via the U.S. CPI, and finally convert them into euros (EUR-2017). Formally, the second step can be written as follows Let $x_{it}$ be an amount of money expressed in terms of currency $i$ at time $t$, and $\eta_{i,\$,t}$ is the exchange rate of currency $i$ per unit of USD at time $t$. $x_{it}$ is converted in $x_{\euro}$ euros at year 2017 as follows:
$$x_{\euro}= \biggl (\frac{x_{it}}{\eta_{i,\$,t}} \times \frac{CPI_{\$,2017}}{CPI_{\$,t}} \biggl)\times \eta_{\euro,\$,2017}$$

When $\eta_{i\$,t}$ is missing, we use a CPI-assisted bridge year $t_0$ (the
closest year to $t$ with an available rate), first re-expressing the amount in
local currency at $t_0$ and then proceeding as above:
$$
x_{\euro} = \left( \left( x_{it} \times \frac{CPI_{i,t_0}}{CPI_{i,t}} \right) \times \frac{1}{\eta_{i,\$,t_0}} \times \frac{CPI_{\$,2017}}{CPI_{\$,t_0}} \right) \times \eta_{\euro,\$,2017}
$$
\noindent where again $t_0$ is the first year closest (to $t$) in which $\eta_{i,\$,t_0}$ is available.

This conversion strategy deserves two brief comments. First using USD as a transit unit is a data-coverage choice since final results are in EUR-2017 (and PPP), so the hub currency is neutral ex post. USD simply offers the broadest, most continuous foreign exchange/CPI backbone. An alternative would be the pound sterling; however, its bilateral coverage against certain CEE and extra-EU currencies is typically thinner than USD. Moreover the dollar's role as a reserve currency yields wide bilateral reporting and fewer idiosyncratic breaks. Therefore, for our goal - maximising coverage and consistency before the final conversion to EUR-2017 - USD is the safer hub. Second, an alternative to converting amounts into USD year-by-year is a CPI-only staging: re-express each amount in constant local currency for a chosen reference year and convert once into a common currency. This approach was used by \cite{Trevisan_EtAl_2011} and \cite{Brugiavini_EtAl_2013} for a subset of countries in SHARELIFE, Wave 3. It has clear advantages -only one currency conversion, and greater robustness in periods of hyperinflation or when official exchange rates diverge from market rates (e.g., in parts of the former Soviet bloc). However, it becomes challenging when the number of currencies is large: CPI series may be unavailable, may not cover all years, or may not exist for defunct currencies. In those cases, a USD-hub workflow with documented CPI assistance used only for some periods and countries might offer more complete coverage.

The third step of our strategy start by improving comparability across countries by proceeding with a Purchasing Power Parity (PPP) adjustment. Particularly, after bringing all amounts to EUR-2017, we exploit PPP-adjusted values to achieve genuine cross-country comparability. Market exchange rates reflect capital flows, policy regimes, and risk premia, and can diverge persistently from differences in price levels faced by households - especially for non-tradable goods (housing, local services). Purchasing Power Parity corrects for these cost-of-living gaps by scaling nominal euros with a country-specific price index, so that one PPP-euro represents the same real purchasing power across countries and over time (we use DE-2017 as the base). This is crucial when comparing wages or benefits across countries, constructing donor pools for imputation, or summarising distributions, because it prevents spurious cross-country gaps driven by exchange-rate volatility rather than real living standards. We still report nominal EUR-2017 for transparency, but PPP-adjusted figures are our preferred scale for cross-national analysis. After the PPP-adjustment, the third step concludes by applying a light two-sided trim on PPP-scaled amounts to curb the influence of outliers and input errors. By default we use the 2.5th-97.5th percentiles for each monetary variable, pooled across countries.

\subsection{Data sources and country-year coverage}
\label{sec:money_data}

We assembled historical exchange rates and CPIs from a small set of long-run, well-documented sources. For bilateral foreign exchange rate levels we combined the International Monteray Fund (IMF) Exchange Rates dataset \citep{imf_er} with MeasuringWorth’s harmonized long-run series by Officer \citep{officer_measuringworth_fx_41} and the Penn World Table–based series distributed via the St. Louis Fed’s FRED portal \citep{fred_pwt_fx}. For specific currency–year windows where national central-bank data are authoritative and higher-frequency, we used the Bank of Russia’s “Dynamics of the Official Exchange Rates” for USD/RUB (series code R01235) \citep{cbr_usd_rub_official_rates_excel}. Consumer prices were sourced from Ljungberg's consolidated European CPIs since 1870 \citep{Ljungberg_2025}, while U.S. CPI levels came from MeasuringWorth \citep{measuringworth_uscpi}.

\begin{figure}[t] 
  \centering
  \includegraphics[width=\linewidth]{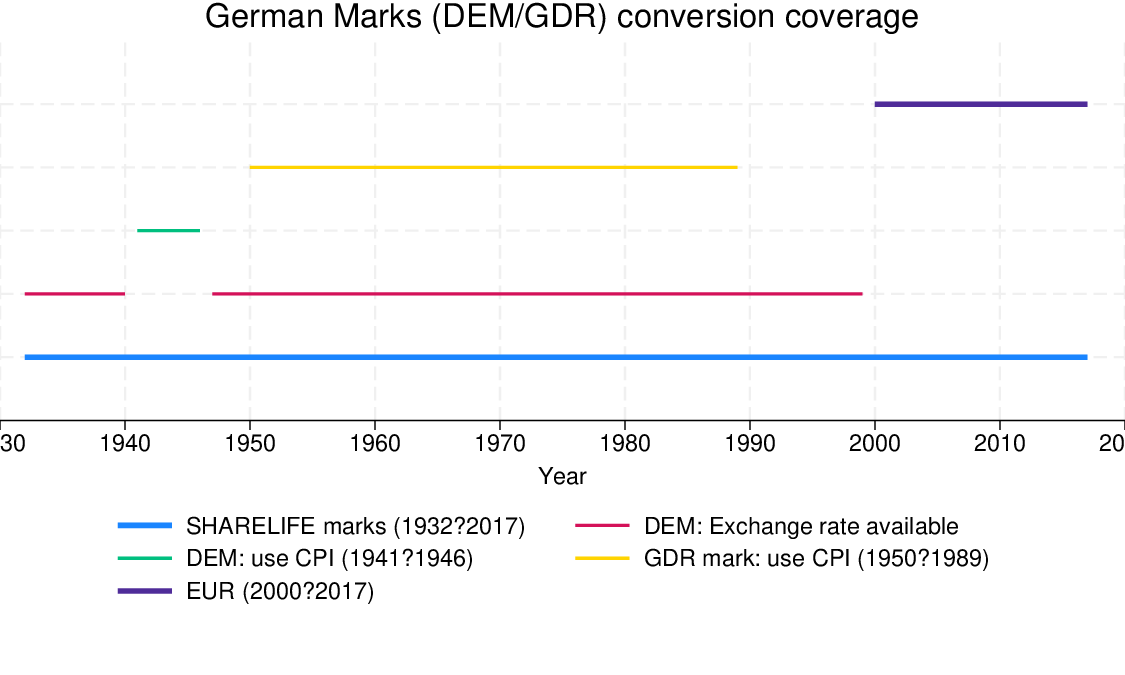}
  \caption{German Marks (DEM/GDR) conversion coverage.}
  \label{fig:mark_coverage}
\end{figure}

\begin{figure}[t] 
  \centering
  \includegraphics[width=\linewidth]{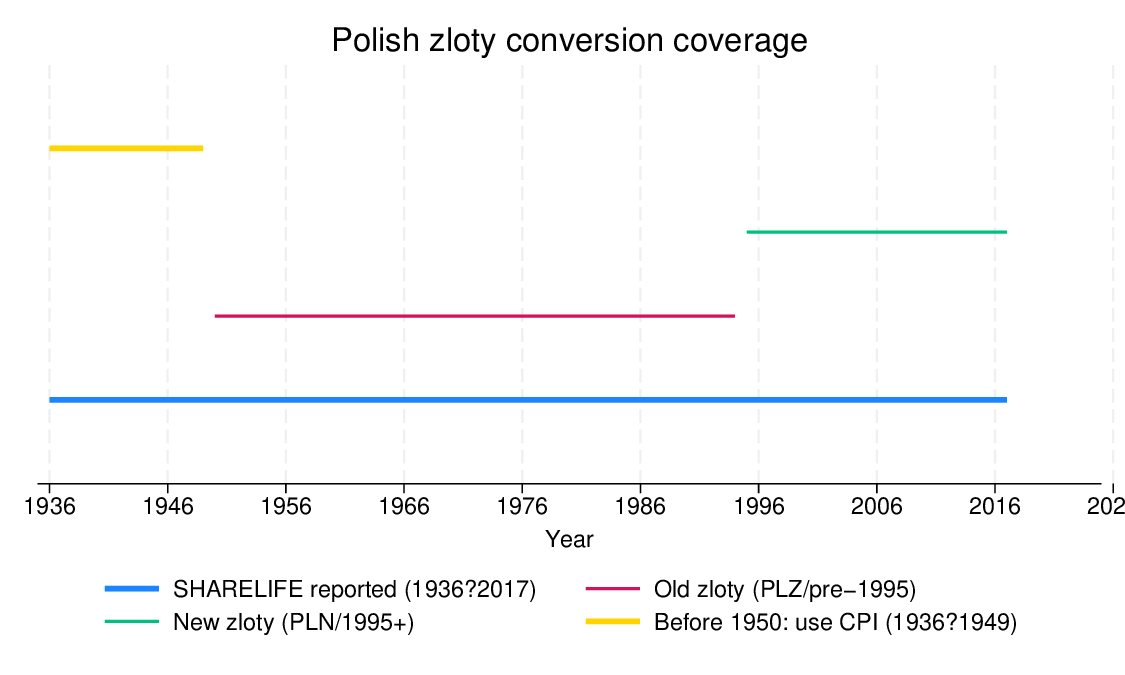}
  \caption{German Marks (DEM/GDR) conversion coverage.}
  \label{fig:zloty_coverage}
\end{figure}

Figures~\ref{fig:mark_coverage} and~\ref{fig:zloty_coverage} summarize the coverage and the conversion strategy for two currencies as an example. In SHARELIFE, amounts reported in złoty—old, new, or unspecified—span 1936-2017. In our pipeline, the ``old'' złoty (PLZ) has bilateral USD exchange rate coverage from 1950 to 1994, while the ``new'' złoty (PLN) is used from 1995 onward following the 1995 redenomination (10,000 PLZ = 1 PLN). For pre-1950 amounts (1936–1949), where USD exchange rates are unavailable, we index values using the Polish CPI \citep{Ljungberg_2025} and link them to the first exchange-rate-covered year before proceeding as usual. The resulting timeline - PLZ 1950-1994, PLN 1995-2017, CPI bridging 1936-1949 - is shown in Figure \ref{fig:zloty_coverage}.

For the German mark, SHARELIFE reports amounts for 1932-2017. Exchange rates for the (West) German mark (DEM) are available in two blocks - 1932-1940 and 1947-1999 - while the WWII gap (1941-1946) is handled via the German CPI \citep{Ljungberg_2025}. For the GDR mark, we use CPI throughout 1950-1989 (no reliable bilateral USD rates), after which values move into the euro period; for completeness we show the EUR span for 2000–2017. Figure \ref{fig:mark_coverage} reports these intervals and bridges.

Operationally, we convert each amount to its USD equivalent using the available exchange rate for that year. When an exchange rate is missing within a documented gap, we first revalue in local real terms with the country-appropriate CPI to the nearest boundary year with an exchange rate, and then continue the standard conversion. Redenominations (e.g., from PLZ to PLN in 1995) are applied before the exchange rate step. All flows are ultimately expressed in a common hub currency (USD) and deflated to a common price year before final conversion to EUR-2017, ensuring consistent treatment across currencies and historical regimes.

To assess conversion coverage, after excluding observations with incomplete information (missing currency type, amount, or year), we retain 242,911 self-reported amounts spanning 148 distinct currencies. Overall, 94.69\% were successfully converted, whereas 12,903 (5.31\%) could not be converted due to uncovered exchange rate, availability of CPI, or inconsistency in the self-reported year/currency (e.g. interviews reporting amounts of euros in an year where euro does not exist as currency). Particularly among the non-converted cases, inconsistencies total 9,417 self-reported amount, accounting for 72.98\% of all non-converted. A glance at Table \ref{tab:convrate} reveals that coverage is dominated by Job-spell first monthly wage, which alone contributes 55.19\% of all records and 55.68\% of converted ones. Other large contributors are Main-job wage (15.16\% to 14.75\%), Pension benefit (13.07\% to 12.70\%), and Current monthly wage (7.63\% to 7.92\%). Conversion rates are generally high—Current main income and Current monthly wage exceed 98\%, while the lowest is Main-job income at 88.63\%. Overall, the results indicate strong conversion coverage, with robustness driven primarily by the wage and pension categories that together account for over 90\% of the data.

Table \ref{tab:convratebycountry} provides the conversion rate by country of interview. Overall conversion is exceptionally high in the Nordics and Switzerland—Sweden (about 99.8\%), Denmark (about 99.8\%), and Switzerland (about 99.5\%) - and similarly strong in Hungary (about 99.7\%) and Bulgaria (about 99.5\%). Although coverage rate is still high, it drops in Croatia (84.0\%), Slovenia (84.7\%), Spain (86.1\%), Romania (86.7\%), and Israel (88.9\%). These drops are mainly due to inconsistency in the year/currency combinations or dominated to ``generic'' currency  labels which cannot be solved using only interviewed country of residence.

\begin{table}[htp]
\caption{Countries for which the CPI has been used}
\begin{center}
\begin{tabular}{lll}
\hline
\multicolumn{1}{c}{SHARELIFE code}  &
\multicolumn{1}{c}{Years}  & \multicolumn{1}{c}{Currency name}  \\
\hline
2   & 1939--1952 & austrian schillings \\
5   & 1941--1944 & belgian francs \\
7   & $<$1952    & bulgarian leva \\
14  & 1942--1945 & swiss francs \\
17  & $<$1989    & german mark (GDR) \\
18  & 1942--1944 & german mark (FRG) \\
19  & 1941--1945 & danish kroner \\
21  & 1942--1945 & spanish peseta \\
22  & 1942--1949 & finnish markkaa \\
25  & 1941--1981 & greek drachmas \\
26  & $<$1951    & irish pounds \\
28  & 1942--1945 & italian lire \\
31  & 1941--1944 & netherlands guilders \\
32  & $<$1950    & polish zloty \\
34  & 1942--1945 & portuguese escudos \\
83  & $<$1993    & czechoslovak koruny \\
111 & $<$1968    & hungarian forints \\
164 & 1941--1945 & norwegian kroner \\
176 & $<$1957    & romanian lei \\
186 & 1942--1945 & swedish kronor \\
\hline
\end{tabular}
\end{center}
\label{tab:CPI}
\end{table}

\begin{table}[htp]
\caption{Conversion coverage by variable}
\begin{center}
\begin{tabular}{lrrr}
\hline
\multicolumn{1}{l}{Variable} &
\multicolumn{1}{c}{Total observations} &
\multicolumn{1}{c}{Converted observations} &
\multicolumn{1}{c}{Conversion rate (\%)}  \\
\hline
$Y_1^{}$   & 11{,}146  & 10{,}739  & 96.35 \\
$Y_2^{}$   & 134{,}060 & 128{,}063 & 95.53 \\
$Y_3^{}$   &  5{,}926  &  5{,}501  & 92.83 \\
$Y_4^{}$   & 31{,}747  & 29{,}213  & 92.02 \\
$Y_5^{}$   & 18{,}532  & 18{,}211  & 98.27 \\
$Y_6^{}$    &  2{,}218  &  2{,}178  & 98.20 \\
$Y_7^{}$           & 36{,}837  & 33{,}933  & 92.12 \\
$Y_8^{}$  &  2{,}409  &  2{,}135  & 88.63 \\
\hline
\end{tabular}
\vspace{-0.7cm}
\parbox{134mm}{\footnotesize
Note: $Y_1^{}$ denotes monthly maternity benefits;
$Y_2^{}$ denotes first monthly wage;
$Y_3^{}$ denotes first monthly income from self-employed;
$Y_4^{}$ denotes monthly pension benefits;
$Y_5^{}$ denotes current monthly wage
$Y_6^{}$ denotes current monthly income
$Y_7^{}$ denotes monthly wage at the end of the main job
$Y_8^{}$ denotes monthly income from self-employment at the end of the main job.}
\end{center}
\label{tab:convrate}
\end{table}

\begin{table}[htp]
\caption{Conversion coverage by country}
\begin{center}
\begin{tabular}{lrrr}
\hline
\multicolumn{1}{l}{Country} &
\multicolumn{1}{c}{Total observations} &
\multicolumn{1}{c}{Converted observations} &
\multicolumn{1}{c}{Conversion rate (\%)}  \\
\hline
AT	&	11,696	&	11,105	&	94.95	\\
BE	&	15,751	&	14,577	&	92.55	\\
BG	&	6,963	&	6,931	&	99.54	\\
CH	&	11,668	&	11,606	&	99.47	\\
CY	&	2,519	&	2,411	&	95.71	\\
CZ	&	19,760	&	19,409	&	98.22	\\
DE	&	16,968	&	16,032	&	94.48	\\
DK	&	12,834	&	12,806	&	99.78	\\
EE	&	15,871	&	15,174	&	95.61	\\
ES	&	8,795	&	7,577	&	86.15	\\
FI	&	5,571	&	5,318	&	95.46	\\
FR	&	11,123	&	9,970	&	89.63	\\
GR	&	4,736	&	4,414	&	93.20	\\
HR	&	5,026	&	4,221	&	83.98	\\
HU	&	5,379	&	5,362	&	99.68	\\
IE	&	2,310	&	2,237	&	96.84	\\
IL	&	3,365	&	2,991	&	88.89	\\
IT	&	12,596	&	11,960	&	94.95	\\
LT	&	7,065	&	6,661	&	94.28	\\
LU	&	2,931	&	2,629	&	89.70	\\
LV	&	5,333	&	4,799	&	89.99	\\
MT	&	2,672	&	2,534	&	94.84	\\
NL	&	6,900	&	6,649	&	96.36	\\
PL	&	12,113	&	11,698	&	96.57	\\
PT	&	2,790	&	2,582	&	92.54	\\
RO	&	4,853	&	4,206	&	86.67	\\
SE	&	13,158	&	13,135	&	99.83	\\
SI	&	5,089	&	4,312	&	84.73	\\
SK	&	7,076	&	6,702	&	94.71	\\
\hline
\end{tabular}
\end{center}
\label{tab:convratebycountry}
\end{table}
\clearpage

\section{Descriptive statistics}
\label{sec:descriptive}

Descriptive statistics on response rates and missing-data patterns should always be examined as a first step.
First, when a variable exhibits a large proportion of missing observations, the imputed values exert greater influence on the results, making careful specification of the imputation model particularly important.
Second, variables displaying a monotone missing-data pattern can be imputed using a sequence of univariate imputation methods.

After removing observations with inconsistent responses, the final database comprises 91,680 records.
Of these, 21,188 (about 23\%) exhibit a complete response pattern with respect to the variables flagged for imputation.

\begin{table}[htp]
\caption{Response rate for some main variables}
\begin{center}
\begin{tabular}{p{10cm}rrr}
\hline
\multicolumn{1}{c}{Country} &
\multicolumn{1}{c}{$E$} &
\multicolumn{1}{c}{$R$} &
\multicolumn{1}{c}{$RR$}  \\
\hline
Age at interview year                                                 &  91,680 &  91,675 &  1.00\\
ISCED coding of education                                             &  91,680 &  90,248 &  0.98\\
Age at end of full-time education                                     &  90,113 &  81,150 &  0.90\\
Number of natural children                                            &  91,680 &  91,372 &  1.00\\
Nt.child 1: Age at childbirth                                         &  80,789 &  80,140 &  0.99\\
Nt.child 2: Age at childbirth                                         &  65,382 &  64,784 &  0.99\\
Nt.child 3: Age at childbirth                                         &  27,632 &  27,147 &  0.98\\
Number of partner                                                     &  91,680 &  91,258 &  1.00\\
Number of non cohabitant partners                                     &  91,680 &  91,175 &  0.99\\
Number of accommodation spells                                        &  91,680 &  91,457 &  1.00\\
Main job in career: Job number                                        &  60,850 &  59,926 &  0.98\\
Number of job spells                                                  &  85,507 &  84,921 &  0.99\\
Job spell 1: Age when job started                                     &  85,507 &  78,094 &  0.91\\
Job spell 2: Age when job started                                     &  58,532 &  55,705 &  0.95\\
Job spell 3: Age when job started                                     &  36,427 &  34,487 &  0.95\\
Job spell 4: Age when job started                                     &  21,715 &  20,335 &  0.94\\
Job spell 5: Age when job started                                     &  12,659 &  11,631 &  0.92\\
Ever had any stocks or shares                                         &  91,680 &  90,664 &  0.99\\
Age when invested in stocks first                                     &  20,505 &  16,899 &  0.82\\
Ever had any mutual funds                                             &  91,680 &  90,495 &  0.99\\
Age invested in mutual funds                                          &  16,729 &  13,389 &  0.80\\
Ever taken out a life insurance policy                                &  91,680 &  90,621 &  0.99\\
Age when taken out a life insurance policy first                      &  29,424 &  23,475 &  0.80\\
Ever owned business                                                   &  91,680 &  91,062 &  0.99\\
Age when first owned business                                         &   3,885 &   3,083 &  0.79\\
Ever had physical injury to disability                                &  91,680 &  91,194 &  0.99\\
When received this injury                                             &  10,357 &   9,566 &  0.92\\
Number periods of ill health                                          &  91,680 &  91,089 &  0.99\\
Illness spell 1: Age at starting year                                 &  16,795 &  14,040 &  0.84\\
Vaccinations during childhood                                         &  91,680 &  89,896 &  0.98\\
Number of times dispossessed because of persecution                   &  91,680 &  90,569 &  0.99\\
\hline
\end{tabular}
\end{center}
\label{tab:des-all}
\end{table}

Table~\ref{tab:des-all} reports the number of eligible respondents ($E$), the number of complete observations ($R$), and the response rate ($RR = R/E$)
for a selection of variables.
Overall completeness is high for basic items, but lower for variables that require recalling precise timing (e.g., the age at which the respondent first invested in mutual funds).

Monetary variables require some special attention because their nonresponse rates can be relatively high, driven by (i) unreported amounts, (ii) currency-conversion issues, and (iii) extreme or implausible values addressed through trimming. As mentioned in the previous section we focus on the following monetary variables:
\begin{itemize}
\item Maternity benefits for natural children (spell data - only women);
\item First monthly wages/income from work (spell data);
\item Pension benefits when retired (spell data);
\item Current monthly wage/income from work (individual data);
\item Wage/income from work at the end of main job (individual data);
\end{itemize}
For these variables, we symmetrically trim 2.5\% of observations from the lower and upper tails of the distribution of each PPP-adjusted variable.

Tables \ref{tab:des-mbn}–\ref{tab:des-iej} report country-level response rates for each monetary variable. The columns in each table are as follows:
\begin{itemize}
\item $E$ - Respondents who are eligible to answer question about a given monetary variable and report a positive value
\item $R_1$ - Respondents who answered the amount question
\item $R_2$ - Respondents for which the amount can be properly converted in $\euro$
\item $R_3$ - Respondents with a PPP-adjusted amount that falls within the trimming interval.
This represents the number of donors that will be available for imputing the missing values.
\item Response/conversion rates:
$$
RR_j=\frac{R_j}{E} \qquad (j=1,2,3)
$$
\end{itemize}
Notice that by construction we have: $RR_1 \ge RR_2 \ge RR_3$.

\begin{table}[htp]
\caption{Response rates for monthly maternity benefits of natural children}
\begin{center}
\begin{tabular}{lrrrrrrrrr}
\hline
\multicolumn{1}{c}{Country} &
\multicolumn{1}{c}{$E$} &
\multicolumn{1}{c}{$R_1$} &
\multicolumn{1}{c}{$RR_1$} &
\multicolumn{1}{c}{$R_2$} &
\multicolumn{1}{c}{$RR_2$} &
\multicolumn{1}{c}{$R_3$} &
\multicolumn{1}{c}{$RR_3$} \\
\hline
AT  &   1,435             &    732 &  0.51         &    683 &  0.48         &    681 &  0.47\\
DE  &   1,317             &    599 &  0.45         &    573 &  0.44         &    565 &  0.43\\
SE  &   2,236             &    454 &  0.20         &    447 &  0.20         &    397 &  0.18\\
NL  &      98             &     43 &  0.44         &     42 &  0.43         &     39 &  0.40\\
ES  &     150             &     37 &  0.25         &     31 &  0.21         &     29 &  0.19\\
IT  &     376             &    161 &  0.43         &    157 &  0.42         &    121 &  0.32\\
FR  &   1,772             &    146 &  0.08         &    137 &  0.08         &    124 &  0.07\\
\hline
\end{tabular}
\end{center}
\label{tab:des-mbn}
\end{table}

\begin{table}[htp]
\caption{Response rates for first monthly wages}
\begin{center}
\begin{tabular}{lrrrrrrrrr}
\hline
\multicolumn{1}{c}{Country} &
\multicolumn{1}{c}{$E$} &
\multicolumn{1}{c}{$R_1$} &
\multicolumn{1}{c}{$RR_1$} &
\multicolumn{1}{c}{$R_2$} &
\multicolumn{1}{c}{$RR_2$} &
\multicolumn{1}{c}{$R_3$} &
\multicolumn{1}{c}{$RR_3$} \\
\hline
AT  &   8,620             &   6,840 &  0.79         &   5,972 &  0.69         &   5,774 &  0.67\\
DE  &  12,646             &  10,830 &  0.86         &   9,556 &  0.76         &   9,326 &  0.74\\
SE  &  14,244             &   8,245 &  0.58         &   7,576 &  0.53         &   7,014 &  0.49\\
NL  &   6,690             &   4,864 &  0.73         &   4,498 &  0.67         &   4,433 &  0.66\\
ES  &   9,041             &   4,929 &  0.55         &   3,983 &  0.44         &   3,552 &  0.39\\
IT  &   8,174             &   6,487 &  0.79         &   5,907 &  0.72         &   5,559 &  0.68\\
FR  &  11,511             &   6,372 &  0.55         &   5,439 &  0.47         &   4,783 &  0.42\\
\hline
\end{tabular}
\end{center}
\label{tab:des-fmw}
\end{table}
\clearpage

\begin{table}[htp]
\caption{Response rates for first monthly incomes from self-employment}
\begin{center}
\begin{tabular}{lrrrrrrrrr}
\hline
\multicolumn{1}{c}{Country} &
\multicolumn{1}{c}{$E$} &
\multicolumn{1}{c}{$R_1$} &
\multicolumn{1}{c}{$RR_1$} &
\multicolumn{1}{c}{$R_2$} &
\multicolumn{1}{c}{$RR_2$} &
\multicolumn{1}{c}{$R_3$} &
\multicolumn{1}{c}{$RR_3$} \\
\hline
AT  &    671             &    341 &  0.51         &    295 &  0.44         &    292 &  0.44\\
DE  &    698             &    492 &  0.70         &    441 &  0.63         &    430 &  0.62\\
SE  &    690             &    348 &  0.50         &    326 &  0.47         &    316 &  0.46\\
NL  &    375             &    192 &  0.51         &    173 &  0.46         &    171 &  0.46\\
ES  &  1,293             &    358 &  0.28         &    308 &  0.24         &    289 &  0.22\\
IT  &  1,292             &    816 &  0.63         &    740 &  0.57         &    698 &  0.54\\
FR  &    892             &    304 &  0.34         &    261 &  0.29         &    245 &  0.27\\
\hline
\end{tabular}
\end{center}
\label{tab:des-fmi}
\end{table}
\clearpage

\begin{table}[htp]
\caption{Response rates for monthly pension benefits when retired}
\begin{center}
\begin{tabular}{lrrrrrrrrr}
\hline
\multicolumn{1}{c}{Country} &
\multicolumn{1}{c}{$E$} &
\multicolumn{1}{c}{$R_1$} &
\multicolumn{1}{c}{$RR_1$} &
\multicolumn{1}{c}{$R_2$} &
\multicolumn{1}{c}{$RR_2$} &
\multicolumn{1}{c}{$R_3$} &
\multicolumn{1}{c}{$RR_3$} \\
\hline
AT  &   2,313             &  2,002 &  0.87         &  1,714 &  0.74         &  1,694 &  0.73\\
DE  &   2,207             &  1,941 &  0.88         &  1,747 &  0.79         &  1,700 &  0.77\\
SE  &   2,500             &  1,762 &  0.70         &  1,758 &  0.70         &  1,708 &  0.68\\
NL  &     648             &    515 &  0.79         &    446 &  0.69         &    413 &  0.64\\
ES  &   2,234             &  1,340 &  0.60         &  1,114 &  0.50         &  1,057 &  0.47\\
IT  &   2,262             &  1,907 &  0.84         &  1,760 &  0.78         &  1,684 &  0.74\\
FR  &   2,442             &  1,741 &  0.71         &  1,474 &  0.60         &  1,404 &  0.57\\
\hline
\end{tabular}
\end{center}
\label{tab:des-pb}
\end{table}
\clearpage

\begin{table}[htp]
\caption{Response rates for current monthly wages}
\begin{center}
\begin{tabular}{lrrrrrrrrr}
\hline
\multicolumn{1}{c}{Country} &
\multicolumn{1}{c}{$E$} &
\multicolumn{1}{c}{$R_1$} &
\multicolumn{1}{c}{$RR_1$} &
\multicolumn{1}{c}{$R_2$} &
\multicolumn{1}{c}{$RR_2$} &
\multicolumn{1}{c}{$R_3$} &
\multicolumn{1}{c}{$RR_3$} \\
\hline
AT  &    436             &    389 &  0.89         &    370 &  0.85         &    359 &  0.82\\
DE  &  1,382             &  1,292 &  0.93         &  1,265 &  0.92         &  1,155 &  0.84\\
SE  &  1,130             &  1,053 &  0.93         &  1,051 &  0.93         &  1,031 &  0.91\\
NL  &    563             &    522 &  0.93         &    495 &  0.88         &    452 &  0.80\\
ES  &    986             &    689 &  0.70         &    590 &  0.60         &    569 &  0.58\\
IT  &    945             &    814 &  0.86         &    806 &  0.85         &    784 &  0.83\\
FR  &  1,052             &    970 &  0.92         &    945 &  0.90         &    883 &  0.84\\
\hline
\end{tabular}
\end{center}
\label{tab:des-cmw}
\end{table}
\clearpage

\begin{table}[htp]
\caption{Response rates for current monthly incomes from self-employment}
\begin{center}
\begin{tabular}{lrrrrrrrrr}
\hline
\multicolumn{1}{c}{Country} &
\multicolumn{1}{c}{$E$} &
\multicolumn{1}{c}{$R_1$} &
\multicolumn{1}{c}{$RR_1$} &
\multicolumn{1}{c}{$R_2$} &
\multicolumn{1}{c}{$RR_2$} &
\multicolumn{1}{c}{$R_3$} &
\multicolumn{1}{c}{$RR_3$} \\
\hline
AT  &    108             &     75 &  0.69         &     71 &  0.66         &     66 &  0.61\\
DE  &    208             &    159 &  0.76         &    157 &  0.75         &    145 &  0.70\\
SE  &    167             &    124 &  0.74         &    124 &  0.74         &    120 &  0.72\\
NL  &     90             &     56 &  0.62         &     54 &  0.60         &     50 &  0.56\\
ES  &    251             &    102 &  0.41         &     90 &  0.36         &     88 &  0.35\\
IT  &    297             &    210 &  0.71         &    208 &  0.70         &    204 &  0.69\\
FR  &    149             &    105 &  0.70         &    103 &  0.69         &     94 &  0.63\\
\hline
\end{tabular}
\end{center}
\label{tab:des-cmi}
\end{table}
\clearpage

\begin{table}[htp]
\caption{Response rates for monthly wages at the end of the main job}
\begin{center}
\begin{tabular}{lrrrrrrrrr}
\hline
\multicolumn{1}{c}{Country} &
\multicolumn{1}{c}{$E$} &
\multicolumn{1}{c}{$R_1$} &
\multicolumn{1}{c}{$RR_1$} &
\multicolumn{1}{c}{$R_2$} &
\multicolumn{1}{c}{$RR_2$} &
\multicolumn{1}{c}{$R_3$} &
\multicolumn{1}{c}{$RR_3$} \\
\hline
AT  &   2,627             &   2,144 &  0.82         &   1,861 &  0.71         &   1,818 &  0.69\\
DE  &   3,008             &   2,475 &  0.82         &   2,201 &  0.73         &   2,115 &  0.70\\
SE  &   2,513             &   1,775 &  0.71         &   1,757 &  0.70         &   1,707 &  0.68\\
NL  &   1,366             &   1,001 &  0.73         &     903 &  0.66         &     882 &  0.65\\
ES  &   2,963             &   1,544 &  0.52         &   1,288 &  0.43         &   1,221 &  0.41\\
IT  &   2,794             &   2,275 &  0.81         &   2,073 &  0.74         &   1,956 &  0.70\\
FR  &   2,755             &   1,730 &  0.63         &   1,494 &  0.54         &   1,384 &  0.50\\
\hline
\end{tabular}
\end{center}
\label{tab:des-wej}
\end{table}
\clearpage

\begin{table}[htp]
\caption{Response rates for monthly incomes from self-employment at the end of main job}
\begin{center}
\begin{tabular}{lrrrrrrrrr}
\hline
\multicolumn{1}{c}{Country} &
\multicolumn{1}{c}{$E$} &
\multicolumn{1}{c}{$R_1$} &
\multicolumn{1}{c}{$RR_1$} &
\multicolumn{1}{c}{$R_2$} &
\multicolumn{1}{c}{$RR_2$} &
\multicolumn{1}{c}{$R_3$} &
\multicolumn{1}{c}{$RR_3$} \\
\hline
AT  &    344             &    171 &  0.50         &    137 &  0.40         &    136 &  0.40\\
DE  &    173             &    117 &  0.68         &     91 &  0.53         &     83 &  0.48\\
SE  &    149             &     96 &  0.64         &     92 &  0.62         &     89 &  0.60\\
NL  &    122             &     55 &  0.45         &     38 &  0.31         &     35 &  0.29\\
ES  &    550             &    189 &  0.34         &    171 &  0.31         &    162 &  0.29\\
IT  &    531             &    345 &  0.65         &    309 &  0.58         &    289 &  0.54\\
FR  &    382             &    152 &  0.40         &    116 &  0.30         &    113 &  0.30\\
\hline
\end{tabular}
\end{center}
\label{tab:des-iej}
\end{table}
\clearpage

Table \ref{tab:des-mbn} describes the response rates for monthly maternity benefit of natural children.
Coverage is modest overall, with sizeable between-country dispersion and very small eligible counts in several countries. The main drop occurs already at the reporting stage ($RR_1$); later losses from conversion and trimming are comparatively minor. This indicator is informative but will benefit from pooling or auxiliary predictors.

Tables~\ref{tab:des-fmw} and~\ref{tab:des-fmi} show the response rates for  first monthly wages and  first monthly incomes from self-employment.
Nonresponse is comparatively high and heterogeneous for first monthly incomes: several countries show low $RR_1$ and further erosion through conversion and trimming, pointing to reporting difficulties and outliers in the tails.
Differently, first monthly wages (employees/civil servants) show a large sample and solid coverage, which make this the backbone of the historical wage evidence.
$RR_1$ is relatively high, while the decline from $RR_2$ to $RR_3$ is moderate, indicating manageable conversion/outlier issues.

Reported amounts for monthly pension benefits when retired in Table~\ref{tab:des-pb}, shows high and stable coverage across most countries: $RR_1$ is close to unity in many cases, and the subsequent drops from conversion and trimming are small.

Current monthly wages and incomes coverage (Tables~\ref{tab:des-cmw} and~\ref{tab:des-cmi}) appears to be lower than for pensions and first wages, with a notable initial shortfall at $RR_1$ and additional losses from conversion/trim in a subset of countries. Contemporary wage reporting appears noisier (possibly due to mixed contract types and currency changes). Regarding current monthly income (self-employment) one observes a mid-to-high coverage overall, with good retention after conversion and trimming in several countries. Remaining gaps are concentrated in a few low-response countries.

Finally response rate for wages and incomes at the end of the main job (see Tables~\ref{tab:des-wej} and \ref{tab:des-iej}) is reasonably strong coverage and large eligible base for the wage, but weaker for income. $RR_1$ declines further through trim, which is consistent with end-of-career spikes and outliers. Results are broadly comparable across countries.
Appendix~B presents scatter plots of the monetary variables against their relevant time dimension (e.g., for maternity benefits, the year of childbirth).

\section{Multiple imputation}
\label{sec:MI}

Multiple imputation (MI) is a popular and flexible simulation-based approach for handling missing data.
It proceeds in two stages: an imputation step and an analysis step.
In the imputation step, missing values are replaced with $M$ independent sets of plausible values drawn from the posterior predictive distribution of the missing data conditional on the observed data.
In the analysis step, the statistical procedure of interest is applied separately to each of the $M$ completed datasets $(m = 1, \ldots, M)$, and the results are subsequently combined into a single MI estimate.
Unlike single imputation methods, which treat imputed values as if they were observed and thus tend to underestimate variability, MI properly accounts for the randomness of the imputed values when estimating standard errors.

Another key feature of the MI approach is the independence of its two steps:
the imputation step fills in the missing values, while the analysis step applies the intended statistical procedures to the completed database without requiring further information about the missing-data mechanism.
This separation makes MI particularly attractive, as it allows the imputation and analysis to be conducted independently, for example, by a data imputer and a data analyst.

In this study, we adopt the perspective of the data producer, aiming to design an imputation model that preserves the key features of the SHARELIFE interview while remaining general enough to support a wide range of analyses on the completed datasets. Foundational contributions by \cite{Rubin_1987}, \cite{Meng_1994}, \cite{Schafer_1997}, and \citet{White_EtAl_2011}, among others, offer valuable guidance on both theoretical and practical aspects of imputation modeling. After briefly reviewing the main approaches and principles, we describe the implementation of the SHARELIFE imputation model.

\subsection{Multivariate imputations}
\label{sec:MI_multivariate}


Two main approaches are commonly employed for multivariate imputation with arbitrary missing-data patterns: joint modeling (JM) and fully conditional specification (FCS).
Under the JM approach, a joint multivariate distribution is assumed for all variables with missing data—most commonly a multivariate normal distribution (see, e.g., \citealt{Schafer_1997}).
Imputed values are then drawn from the corresponding posterior predictive distribution of the missing data, conditional on the observed data.
By contrast, the FCS approach of \citet{VanBuuren_EtAl_1999} does not rely on specifying a joint multivariate distribution.
Instead, it imputes multiple variables iteratively through a sequence of univariate imputation models---one for each incomplete variable---using fully conditional prediction equations that include all other variables as predictors.
A comprehensive comparison between multivariate normal imputation and FCS can be found in \citet{Lee_Carlin_2010}.

The SHARELIFE imputation model is largely based on the FCS approach, which is also known as multivariate imputation by chained equations (MICE; \citealt{Royston_2005b}, \citeyear{Royston_2007}, \citeyear{Royston_2009}; \citet{White_EtAl_2011})
and sequential regression multivariate imputation
(SRMI; \citealt{Raghunathan_EtAl_2001}).
Conceptually, FCS is similar to the Gibbs sampling algorithm (\citealt{Gelfand_Smith_1990}), a Markov chain Monte Carlo method used to simulate complex multivariate distributions.
However, unlike Gibbs sampling, the conditional densities in FCS do not necessarily correspond to any valid joint multivariate distribution---a limitation known as incompatibility of conditionals (see, e.g., \citealt{Arnold_EtAl_1999}, \citeyear{Arnold_EtAl_2001}; \citealt{VanBuuren_2007}).
Despite this lack of full theoretical justification, FCS has become one of the most widely adopted multivariate imputation techniques due to its flexibility in handling complex data structures and multiple types of variables
(continuous, count, binary, ordered and unordered categorical), as well as its ability to preserve relationships and constraints among imputed variables (\citealt{Raghunathan_EtAl_2001}; \citealt{VanBuuren_EtAl_2006}).
Furthermore, FCS is the imputation method used for the monetary variables collected in all regular waves of SHARE (see, e.g., \citealt{Bergmann_EtAl_2024}).

To formalize the basic ideas of the FCS approach, let $Y=(Y_{1}^{}, Y_{2}^{}, \ldots, Y_{J}^{})$ be a set of $J$ incomplete variables affected by missing values, and let $Z$ be a set of fully observed variables.
The values of $Y_{j}^{}$ at iteration $t\ge 0$ are denoted by $Y_j^{(t)}$.
At iteration $t=0$, we initialize the components of $Y$ through monotone imputations.
More precisely, the imputed values of $Y_j^{(0)}$ $(j=1,\ldots,J)$ are obtained from a conditional density of the form
\begin{equation}
Y_{j}^{(0)}\sim f_{j}^{}(Y_{j}^{}|Y_{1}^{(0)}, \ldots, Y_{j-1}^{(0)}, Z; \theta_{j}^{}),
\label{eq:FCS_start}
\end{equation}
where the specification of $f_{j}^{} (\cdot)$ depends on the imputation model for $Y_{j}^{}$ (see Section~\ref{sec:MI_imp_eq}), and
$\theta_{j}^{}$ is the corresponding parameter vector, for which we assume a noninformative uniform prior.
Because the ordering of the variables in $Y$ affects the starting values, it is common practice to sort them from the most to the least observed variable.

At iteration $t\ge1$, we update the imputed values of $Y_{j}^{}$ by drawing from a conditional density of the form
\begin{equation}
Y_{j}^{(t)}\sim f_{j}^{}(Y_{j}^{}|Y_{1}^{(t)}, \ldots, Y_{j-1}^{(t)}, Y_{j+1}^{(t-1)},\ldots, Y_{J}^{(t-1)}, Z; \psi_{j}^{}),
\label{eq:FCS_iter}
\end{equation}
where the conditioning set includes the most recently updated imputations of the other $J-1$ variables in $Y$, and $\psi_{j}^{}$ is the corresponding parameter vector, again assumed to have a noninformative uniform prior.
Note that the variables $Y_{1}^{(t)}, \ldots, Y_{j-1}^{(t)}$
have been imputed $t$ times, while the variables $Y_{j+1}^{(t-1)},\ldots, Y_{J}^{(t-1)}$ have been imputed $t-1$ times.
Once the algorithm is iterated over a predefined burn-in period $(t=1,\ldots,T)$, a final set of imputed values is obtained from the last iteration.
Furthermore, multiple imputations are obtained by repeating the above steps independently $M$ times, using different random seeds.

Although not made explicit for ease of notation, the FCS algorithm can be customized to incorporate equation-specific constraints and to operate within predefined subsamples.
These modifications can be regarded as simplifications of the baseline algorithm and are particularly useful for managing complex data structures such as longitudinal data or retrospective life-history data.
In Appendix~A14, we describe the simplifications applied to the monetary variables collected within SHARELIFE.
Our imputation equations depart from~\eqref{eq:FCS_iter} in several important respects.
First, we assume that the serial correlation in the sequences of maternity benefits, first monthly wages, and first monthly incomes is adequately captured by stationary AR(1) processes, whereby the characteristics of the $h$th spell are assumed to depend only on those of the preceding spell.
Second, following the two-fold FCS approach proposed by \cite{Nevalainen_EtAl_2009} and \cite{Welch_EtAl_2014},
we adopt a chain-of-chains structure involving two nested chains: one running over the spells and the other over the variables.
The main difference is that, to simplify the treatment of retrospective life-history data, we ignore the possible dependence on future events.
Third, we specify fully parametric forms for the functions $f_j(\cdot)$, incorporating equation-specific exclusion restrictions and allowing for potential inter-dependencies among the variables to be imputed.

\subsection{Imputation equations}
\label{sec:MI_imp_eq}

The flexibility of the FCS algorithm stems from the fact that, as with monotone imputations, this method relies on a sequence of univariate imputation equations, which may differ across variables depending on their type or distribution.

Let $y=(y_{1}^{}, \ldots, y_{n}^{})_{}'$ represent the vector of observations for the variable to be imputed in a generic iteration of the FCS algorithm, with $n$ denoting the sample size.
We consider different models depending on the nature of the outcome of interest.
When $y_i$ $(i=1,\ldots,n)$ is a continuous variable with an unrestricted range, it is common to specify a Gaussian homoskedastic linear regression model of the form
\begin{equation}
y_{i}^{}|X_{i}^{}\sim\rN(X_{i}'\beta, \sigma_{}^2),
\label{eq:mod_reg}
\end{equation}
where $X_{i}^{}$ denotes the $i$th observation on a set of $k$ predictors, $\beta$ is the corresponding parameter vector, and $\sigma^2$ is the variance.
Let us partition $y = (y_o', y_m')'$ into two subvectors: the $n_0 \times 1$ subvector $y_o$ of observed (complete) values and the $n_1 \times 1$ subvector $y_m$ of missing values, with $n = n_0 + n_1$.
We apply the same partitioning to the $n \times k$ matrix of predictors $X = (X_o', X_m')'$,
so that $X_o$ has dimension $n_0 \times k$ and $X_m$ has dimension $n_1 \times k$.
Like \cite{Gelman_EtAl_1995b},
we employ the following algorithm to simulate the missing data from the (Bayesian) posterior predictive distribution of $y_m^{}$ conditional on $(y_o^{},X_o^{})$:
\begin{enumerate}
\item Fit model~\eqref{eq:mod_reg} to the observed data $(y_o^{},X_o^{})$ and obtain the least squares (LS) estimate $(\hat \beta_{}',\hat \sigma_{}^{2})'$ of $(\beta_{}',\sigma_{}^{2})_{}'$.

\item
Draw the new values of the parameters $(\beta_{}',\sigma_{}^{2})_{}'$ from their joint posterior distribution under a noninformative improper prior:
$\sigma_{*}^{2}\sim \hat \sigma^2 (n_{0}^{}-k)/ \chi_{n_{0}^{}-k}^2$ and
$\beta_{*}^{}|\sigma_{*}^{2} \sim \rN(\hat \beta,\sigma_{*}^{2} (X_o'X_{o}^{})_{}^{-1})$.

\item Draw the new values of the missing observations $y_m^{}$ from the posterior predictive distribution
$\rN(X_m^{}\beta_{*},\sigma_{*}^2 I_{n_{1}^{}}^{})$, where $I_{n_{1}^{}}^{}$ denotes the identity matrix of order $n_{1}^{}$.

\item Obtain multiple imputations of $y_m$ by independently repeating the previous two steps under different random seeds.

\end{enumerate}

This approach produces reliable results under the assumptions of linearity and homoskedastic Gaussian errors.
These assumptions are sometimes more plausible after applying suitable transformations to the outcome variables, such as the logarithm, inverse hyperbolic sine, or Box–Cox transformations. In general, however, violations of these distributional assumptions may result in implausible or excessively large imputed values.

One approach to mitigate these issues is predictive mean matching (PMM; \citealt{Rubin_1986}; \citealt{Little_1988}), a partly semi-parametric approach that preserves the observed data distribution.
Unlike other imputation methods, PMM does not draw imputed values directly from the posterior predictive distribution. Instead, it samples them from the observed values of $y_o$.
The key difference lies in the third step of the algorithm: for each missing value $y_i^{}$ with covariates $X_i^{}$, PMM identifies a set of $q\ge 1$ observed units whose predicted means are closest to that of $y_i^{}$, that is, those with the smallest values of $|X_i'\beta_{*}^{} - X_j'\hat \beta|$ $(j=1,\ldots,n_0)$.
Finally, one donor is randomly selected from this set, and its observed value is used as the imputed value for $y_i^{}$.

In addition to the Gaussian imputation model and PMM, we consider a latent linear model that accommodates outcomes measured as point data, interval data, or data subject to left- or right-censoring.
This generalization of the Tobit model, known as interval regression
(\citealt{Wooldridge_2016}; \citealt{Cameron_Trivedi_2022}), is particularly useful for ensuring that imputed values remain within unit-specific bounds.
In this case, the first step of the procedure involves estimating the model parameters by maximum likelihood (ML).
In the third step, the imputed values of $y_m$ are then drawn from a truncated normal distribution, thereby guaranteeing that each imputed value falls within its prespecified interval.

For cases where $y_i$ is a count variable, we consider a Poisson model of the form
\begin{equation}
\Pr(y_i^{}=c|X_{i}^{})=e^{-\lambda_i} \lambda_i^c / c!,
\qquad c=0,1,2,\ldots,
\label{eq:mod_poi}
\end{equation}
where $\lambda_i(\beta)=\exp(X_i'\beta)$.
Fitting model~\eqref{eq:mod_poi} to the observed data $(y_{o}^{},X_{o}^{})$, we first obtain the ML estimate $\hat \beta$ of $\beta$ and its estimated variance matrix $\hat V$.
In the second step, we draw $\beta_{*}^{}$ from $\rN(\hat \beta, \hat V)$.
In the third step, we impute the missing values in $y_m^{}$ by simulating from the Poisson distribution~\eqref{eq:mod_poi} with intensity parameter $\lambda_i(\beta_{*}^{})$.

Similarly, when $y_i$ is a binary indicator, we specify a logit model of the form
\begin{equation}
\Pr(y_i^{}=1|X_{i}^{})=[1+\exp(-X_i'\beta)]_{}^{-1}.
\label{eq:mod_log}
\end{equation}
Fitting model~\eqref{eq:mod_log} to the observed data $(y_{o}^{},X_{o}^{})$, we obtain the ML estimate $\hat \beta$ of $\beta$ and its estimated variance matrix $\hat V$.
In the second step, we draw $\beta_{*}^{}$ from $\rN(\hat \beta, \hat V)$.
In the third step, we impute the missing values in $y_m^{}$ by simulating from the logistic distribution~\eqref{eq:mod_log} with $\beta$ replaced by $\beta_{*}^{}$.

The logit imputation model naturally extends to ordered and unordered categorical variables through the use of ordered and multinomial logit specifications, respectively.
However, imputation models for categorical variables may encounter the problem of perfect prediction, which arises when estimated outcome probabilities are exactly 0 or 1 for one or more observations.
Perfect prediction can result in unstable draws of $\beta_{*}^{}$ and unreliable variance estimates.
To mitigate this issue, we adopt the data augmentation technique proposed by \cite{White_EtAl_2010}, which adds a small number of pseudo-observations with minimal weights to the dataset, thereby preventing perfect prediction.

We conclude this section by noting that appropriate combinations of the previously described models can be used to impute semi-continuous variables, that is, variables characterized by a substantial fraction of zeros and a continuous distribution among the positive values.
A common strategy in such cases is the two-part model, which consists of a logit model for predicting whether the outcome is zero or positive, followed by a linear regression model (or PMM) for the strictly positive part of the distribution.

\subsection{Choice of predictors and data pooling strategies}
\label{sec:MI_predictors}


Most of our imputation equations are estimated separately by macro-region and include a full set of country dummies to capture unobserved cross-country heterogeneity within each macro-region.
Based on considerations of sample size and country homogeneity, we distinguish six macro-regions:
Saxons (SAX; Austria, Germany, the Netherlands, and Switzerland),
Nordic (NO; Denmark, Estonia, Finland, Latvia, Lithuania, and Sweden),
Central Europe (CE; Czech Republic, Hungary, and Poland),
Mediterranean (MED; Cyprus, Greece, Italy, Malta, Portugal, and Spain),
Balkans (BAL; Bulgaria, Croatia, Romania, and Slovenia),
and
Mixed (MIX; Belgium, France, Ireland, Israel, and Luxembourg).
Using this partition, the number of respondents ranges from a minimum of 10,211 in the Balkans to a maximum of 19,245 in the Mediterranean region.
In practice, however, the available sample size may differ across variables due to skip patterns in the SHARELIFE questionnaires of wave 3 and wave 7.
When the variable-specific sample size within a macro-region is too small (e.g., fewer than 50 observations), we pool data from multiple macro-regions and replace the country dummies with a corresponding set of macro-region dummies.
This situation typically arises for event sequences containing a large number of spells, where only a limited number of respondents report data for several of them.
In these cases, models for the first spells are estimated separately by macro-region, while those for the later spells are estimated on pooled data to ensure sufficient sample size and stable parameter estimates.

In addition to the variables that are imputed jointly in the FCS method and the country or macro-region fixed effects, our set of predictors typically includes basic socio-demographic characteristics of respondents and features of the interview process.
In particular, we use a second-order polynomial in the respondent’s age at the time of the interview, household size, and a set of binary indicators for being female, living with a spouse or partner, having a high educational level, reporting good health, having no more than ten books at home during childhood, participating in the second SHARELIFE interview, being interviewed with the help of a proxy respondent, and being perceived by the interviewer as having a good understanding of the questions.
To ensure cross-country comparability, educational attainment information has been recoded according to the 1997 International Standard Classification of Education (ISCED-97), using data collected in the regular waves of SHARE.
Data on gender and household composition are fully observed from the CV\_R modules of previous SHARE waves, whereas small fractions of missing values for the other predictors are imputed separately in the initial modules of the imputation procedure.
For further details see Appendix~A.



\subsection{Specification and implementation of the SHARELIFE imputation model}
\label{sec:MI_specification}

The SHARELIFE imputation model is organized into 14 building blocks, comprising a total of 658 incomplete variables to be imputed. Rather than imputing all these variables jointly, we designed a sequence of separate chains corresponding to different modules of the SHARELIFE interview and different sub-samples of the original data.
The imputation model is implemented in Stata version 19.0, allowing end users to leverage the suite of Stata MI commands during the analysis stage.

We first carried out preliminary recoding and standardization of the raw data from SHARELIFE waves 3 and 7. Specifically, a separate Stata do-file was created for each module of the interview. Key features of the imputation model were kept flexible through the use of local and global macros. For example, although the programs are designed to work with the publicly available SHARELIFE data, individual and household identifiers are defined via global macros, which allows the same programs to be easily adapted to the unscrambled version of the SHARELIFE data.
Within each module, we focused on variables present in both waves, omitting variables recorded in only a single wave. For each variable, we identified the set of eligible respondents, converted SHARE-specific missing data codes into Stata’s standard missing codes, and corrected other logical inconsistencies in the observed data. All steps were fully logged, ensuring transparency, reproducibility, and facilitating subsequent review.

The second step involved harmonizing the monetary data, following the procedures described in Section~\ref{sec:money_conv}.
Using PPP-adjusted amounts, we also performed outlier trimming and computed descriptive statistics on response rates and missing data patterns.

Finally, we implemented the core of the SHARELIFE imputation model, focusing on the following blocks of variables:
(i) socio-demographic variables, proxy reporting, and interviewer observations on the quality of the interview process;
(ii) family composition variables;
(iii) childhood circumstances;
(iv) child history;
(v) partner history;
(vi) accommodation history;
(vii) work history;
(viii) work quality history;
(ix) disability benefits;
(x) financial history;
(xi) health history;
(xii) health care;
(xiii) general life; and
(xiv) monetary variables.
For further details on this stage of the SHARELIFE imputation model, we refer the reader to Appendix~A.

\begin{table}[htp]
\caption{Computational burden of the SHARELIFE imputation model}
\begin{center}
\begin{tabular}{lrr}
\hline
\multicolumn{1}{l}{} &
\multicolumn{1}{c}{Number of }  &
\multicolumn{1}{c}{Computing}  \\
\multicolumn{1}{l}{Module} &
\multicolumn{1}{c}{variables}  &
\multicolumn{1}{c}{time (hours)}  \\
\hline
Socio-demographics       &   17  & 0.70 \\
Family composition       &    4  & 0.20 \\
Childhood circumstances  &   11  & 0.59 \\
Child history            &   128  & 4.14 \\
Partner history          &    32  & 1.02 \\
Accommodation history    &   119  & 2.43 \\
Work history             &  152   & 3.81 \\
Work quality history     &   24  & 0.93 \\
Disability benefits      &    8 & 0.22 \\
Financial history        &    8 & 0.48 \\
Health history           &   23 & 0.55 \\
Health care              &   19  & 1.19 \\
General life             &   52 & 1.92 \\
Monetary variables       &    61 & 9.31 \\
\hline
Total                    & 658 &27.26 \\
\hline
\end{tabular}
\end{center}
\label{tab:ctime}
\end{table}

We generate five multiple imputations for the missing values in each variable, following a burn-in period of 10 iterations.
Both the number of imputations and the length of the burn-in period are specified via global macros,
allowing for easy adjustment.
Table~\ref{tab:ctime} summarizes the computational burden.
Using a ThinkPad laptop with an 11th Gen Intel(R) Core(TM) i7-11800H processor, 32 GB of RAM, and Stata/MP 8 (version 19.0),
the SHARELIFE imputation model completes in approximately 27 hours.
Roughly one third of the computing time can be attributed to the imputation of monetary variables,
which relies on two nested chains (see Appendix~A14).

\subsection{The SHARELIFE-MI database}
\label{sec:MI_output}

The SHARELIFE-MI database contains 550,080 observations on 676 variables.
The data are organized in Stata’s \texttt{flong} style for multiple imputations, with the variable \texttt{mergeid} serving as the individual identifier and \texttt{\_mi\_m} indicating the multiple imputation index.
Accordingly, there are 91,680 observations for each value of \texttt{\_mi\_m}, which ranges from 0 to 5 in increments of 1.
Overall, the database includes 3 system variables, 15 regular variables, and 658 imputed variables. All variables are properly registered and include appropriate variable and value labels.
Note that ineligible cases for each variable are coded as \texttt{-99}, indicating `Not eligible'.
The data can be easily transformed into other formats using Stata’s suite of MI commands.


\section{Diagnostics and validations}
\label{sec:validation}

In this section, we assess the internal and external consistency of our multiple imputations for SHARELIFE.
Although diagnostics for the internal validity of multiple and multivariate imputations remain an active and evolving area of research
(see, e.g., \citealt{Gelman_EtAl_2005}; \citealt{Abayomi_EtAl_2008}; \citealt{Marchenko_Eddings_2011}; \citealt{Eddings_Marchenko_2012}),
two general recommendations consistently emerge from the literature:
(i) evaluating how well the specified imputation model fits the observed data, and
(ii) comparing the distributions of imputed and observed values to ensure that the imputed data are both plausible and coherent with the empirical evidence.
Regarding external validity, potential limitations may stem from the comparability of outcomes across different surveys or studies.
For brevity, we focus our attention on the monetary variables, which are characterized by substantially higher rates of nonresponse.

\begin{figure}[htp]
\centering
\includegraphics[width=1.0\linewidth,height=0.40\textheight]{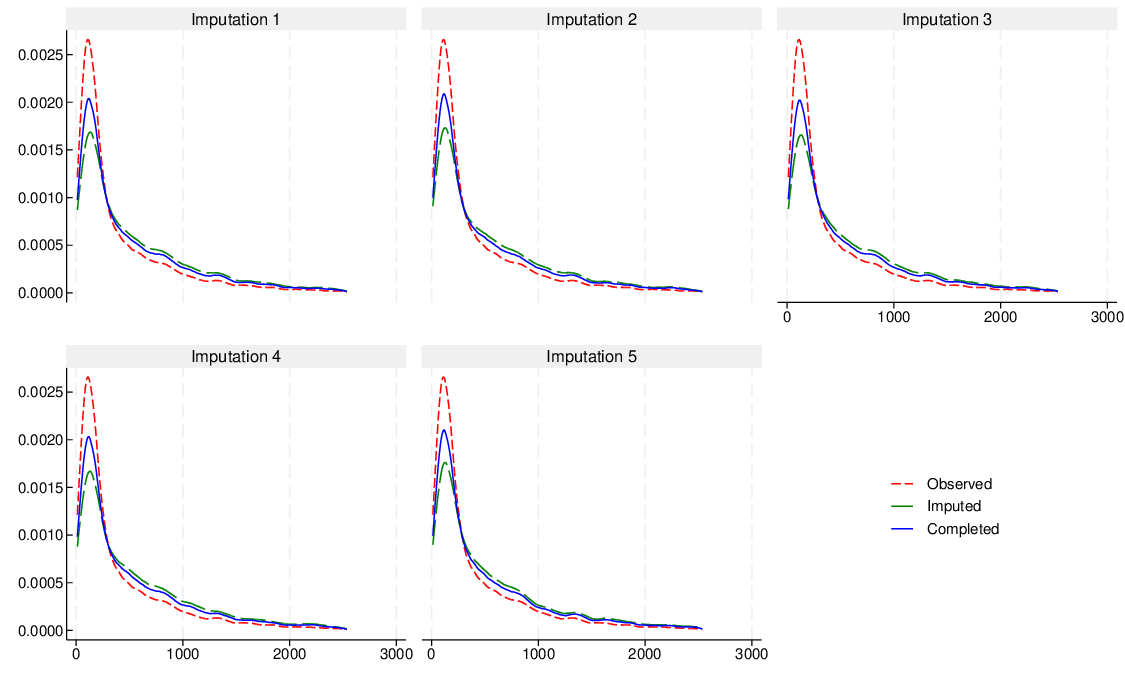}
\caption{Kernel densities of monthly maternity benefits in the observed, imputed, and completed samples by imputation number}
\label{fig:kden_mb}
\end{figure}

\begin{figure}[htb]
\centering
\includegraphics[width=1.0\linewidth,height=0.40\textheight]{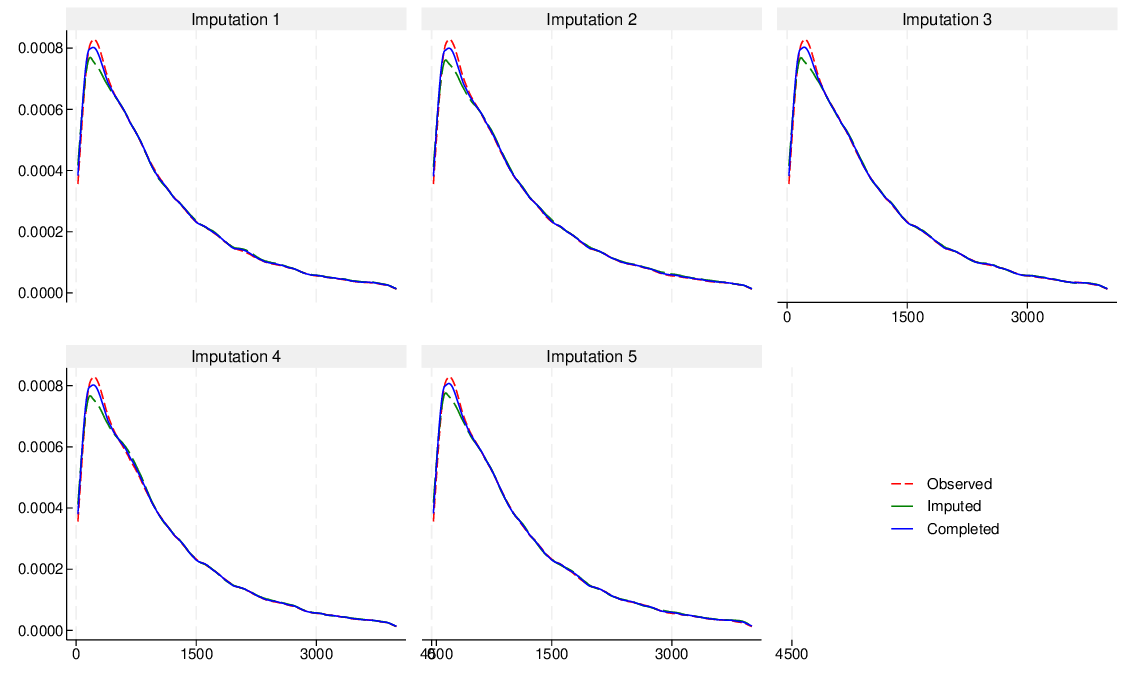}
\caption{Kernel densities of first monthly wages in the observed, imputed, and completed samples by imputation number}
\label{fig:kden_jsp_wage}
\end{figure}

\begin{figure}[htp]
\centering
\includegraphics[width=1.0\linewidth,height=0.40\textheight]{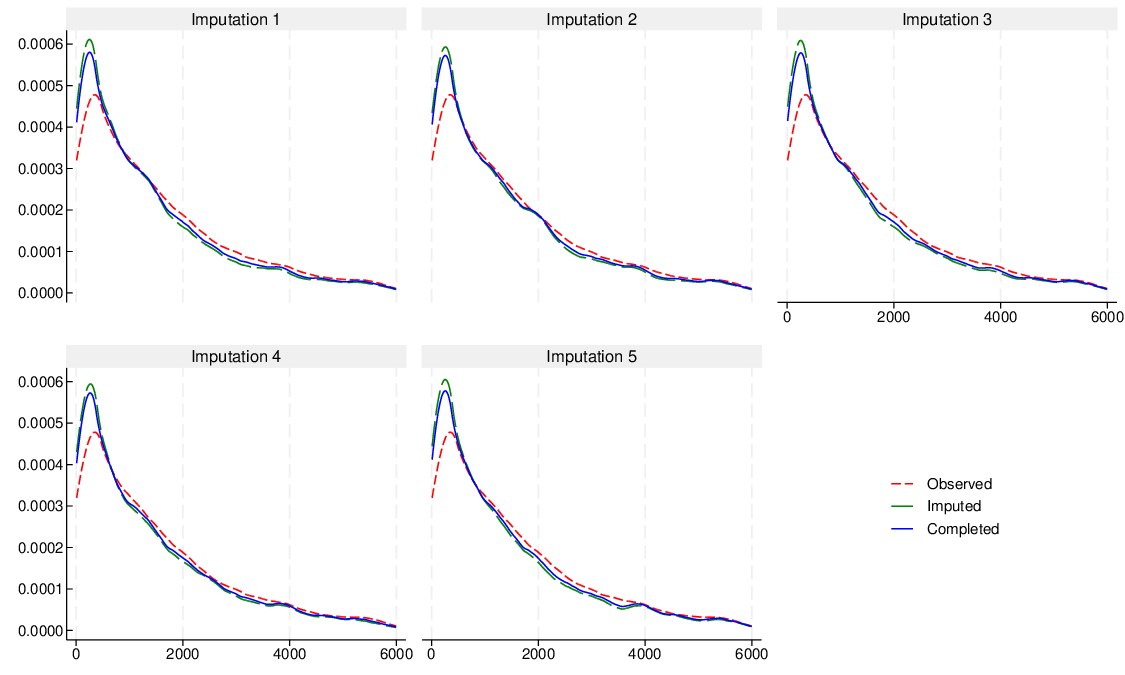}
\caption{Kernel densities of first monthly incomes from self-employment in the observed, imputed, and completed samples by imputation number}
\label{fig:kden_jsp_income}
\end{figure}

\begin{figure}[htp]
\centering
\includegraphics[width=1.0\linewidth,height=0.40\textheight]{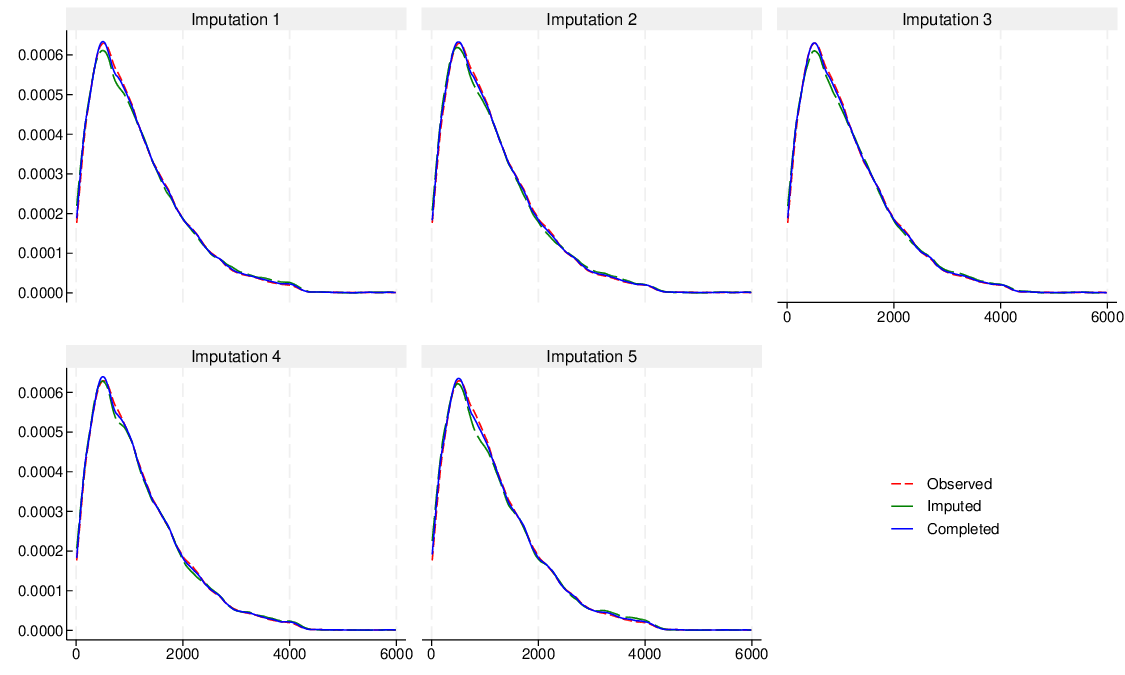}
\caption{Kernel densities of pension benefits when retired in the observed, imputed, and completed samples by imputation number}
\label{fig:kden_pbr}
\end{figure}

\begin{figure}[htp]
\centering
\includegraphics[width=1.0\linewidth,height=0.40\textheight]{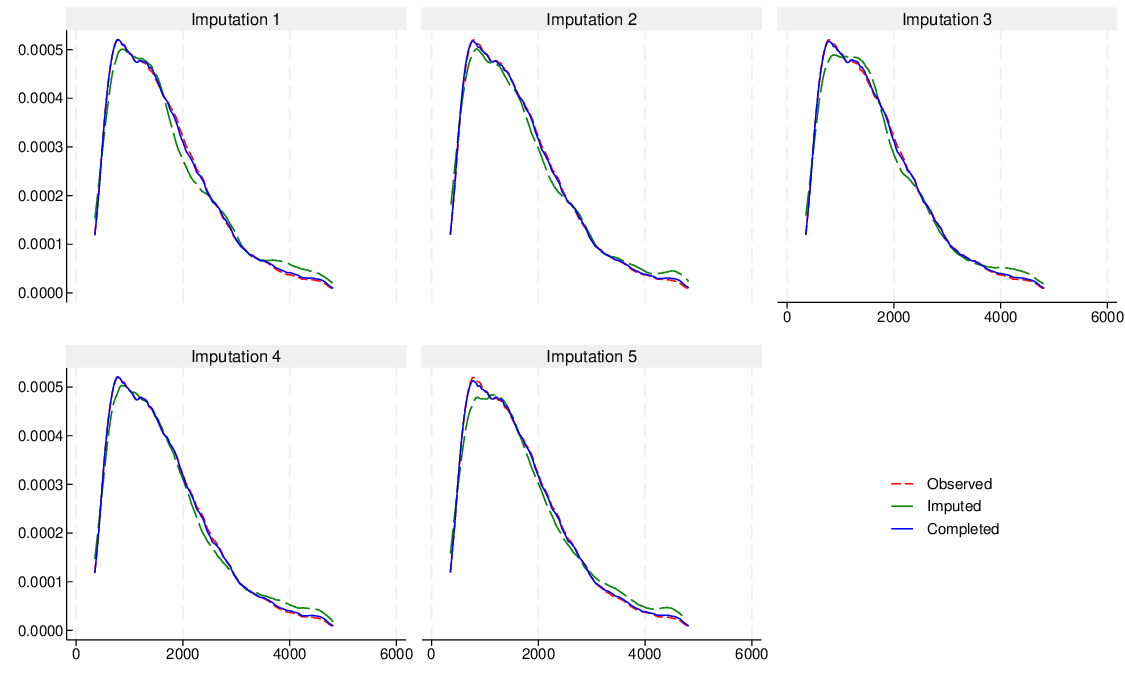}
\caption{Kernel densities of current monthly wages in the observed, imputed, and completed samples by imputation number}
\label{fig:kden_cjob_wage}
\end{figure}

\begin{figure}[htp]
\centering
\includegraphics[width=1.0\linewidth,height=0.40\textheight]{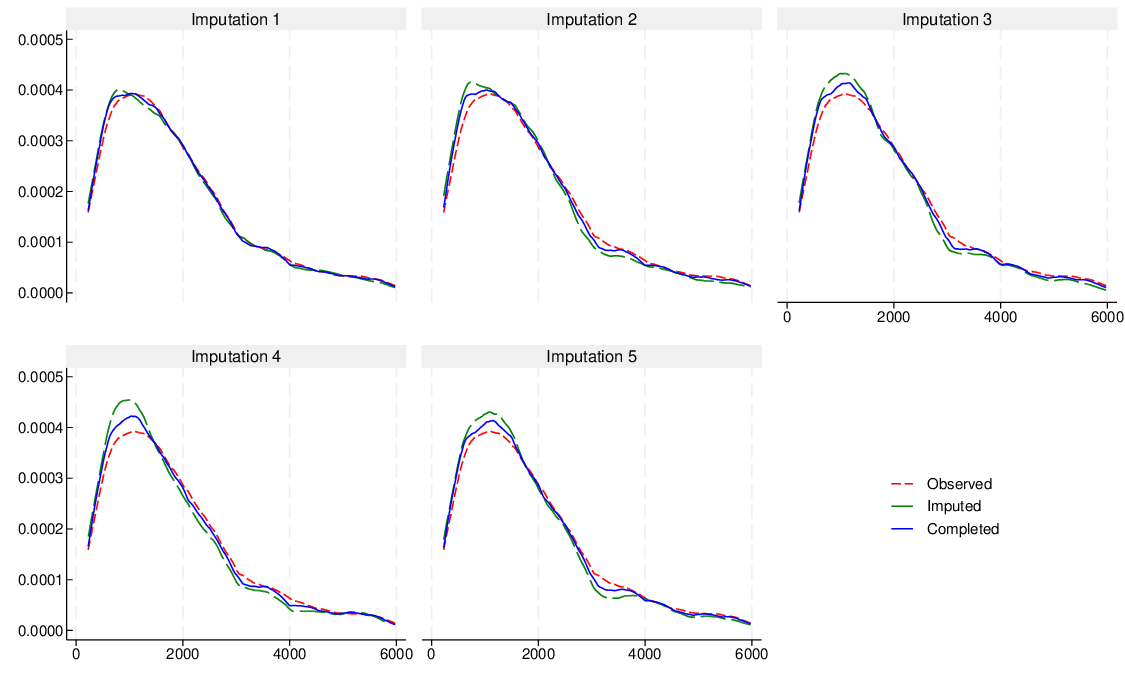}
\caption{Kernel densities of current monthly incomes from self-employment in the observed, imputed, and completed samples by imputation number}
\label{fig:kden_cjob_income}
\end{figure}

\begin{figure}[htp]
\centering
\includegraphics[width=1.0\linewidth,height=0.40\textheight]{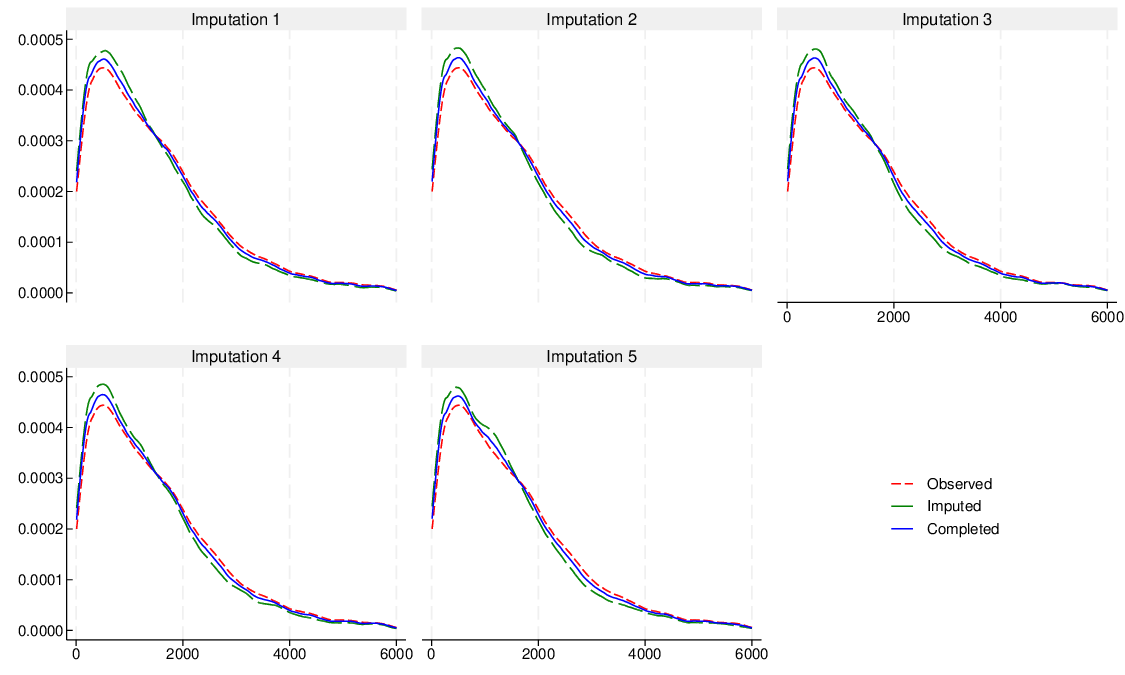}
\caption{Kernel densities of monthly wages at the end of the main job in the observed, imputed, and completed samples by imputation number}
\label{fig:kden_mjob_wage}
\end{figure}

\begin{figure}[htp]
\centering
\includegraphics[width=1.0\linewidth,height=0.40\textheight]{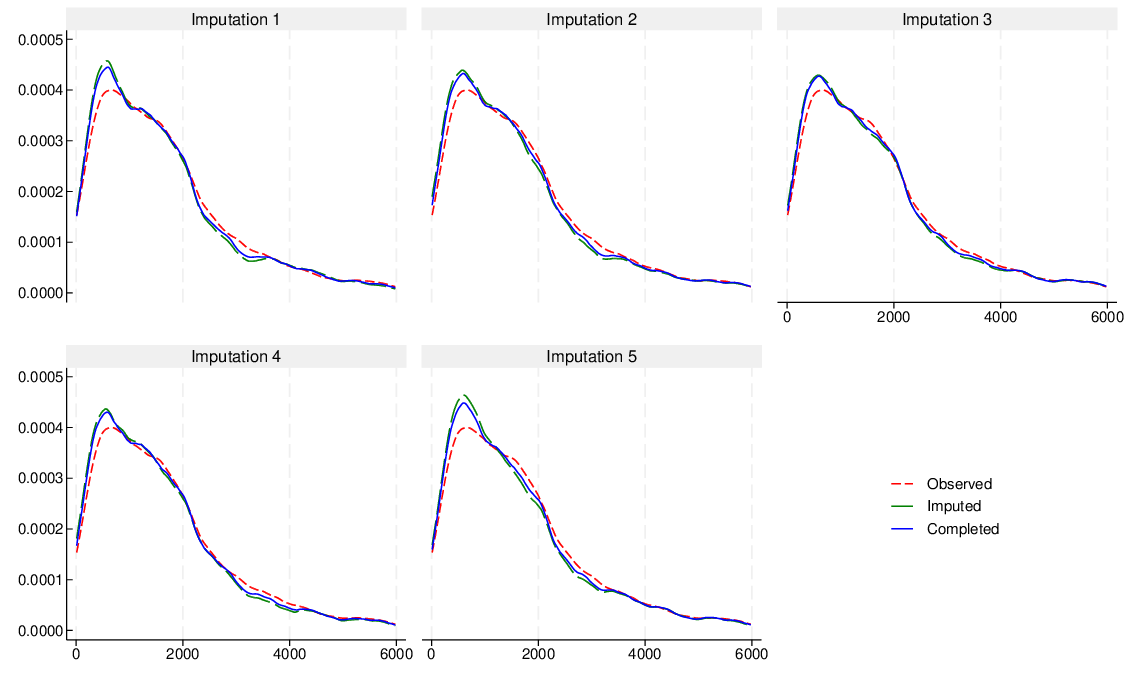}
\caption{Kernel densities of monthly incomes from self-employment at the end of the main job in the observed, imputed, and completed samples by imputation number}
\label{fig:kden_mjob_income}
\end{figure}

Let us first address the issue of internal validity.
Figures~\ref{fig:kden_mb}–\ref{fig:kden_mjob_income} compare the kernel density plots of PPP-adjusted monetary variables across the observed, imputed, and completed (observed plus imputed) samples, disaggregated by imputation number.
Specifically, Figure~\ref{fig:kden_mb} refers to monthly maternity benefits;
Figure~\ref{fig:kden_jsp_wage} to first monthly wages;
Figure~\ref{fig:kden_jsp_income} to first monthly incomes from self-employment;
Figure~\ref{fig:kden_pbr} to monthly pension benefits when retired;
Figure~\ref{fig:kden_cjob_wage} to current monthly wages;
Figure~\ref{fig:kden_cjob_income} to current monthly incomes from self-employment;
Figure~\ref{fig:kden_mjob_wage} to monthly wages at the end of the main job;
and
Figure~\ref{fig:kden_mjob_income} to monthly incomes from self-employment at the end of the main job.
In all cases, the plots display the densities of PPP-adjusted monetary amounts conditional on both eligibility and ownership. For monthly maternity benefits, first monthly wages, and first monthly incomes from self-employment, data from multiple spells are pooled.
Multiple imputations of missing values are plotted across the different panels.

Differences between the observed and imputed distributions of each variable may reflect specific features of the imputation model or the effects of the MAR assumption, but they should remain plausible within the context of the analysis. In our case, the results indicate that the imputation model does not produce implausible values or distort the overall shape of the observed distributions.
Most densities of the imputed values are consistently unimodal and closely aligned with those of the observed data, with only minor variation across imputations.
The observed and imputed distributions of first monthly wages (Figure~\ref{fig:kden_jsp_wage}), monthly pension benefits (Figure~\ref{fig:kden_pbr}), current monthly wages (Figure~\ref{fig:kden_cjob_wage}) and current monthly incomes (Figure~\ref{fig:kden_cjob_income}), as well as monthly wages and incomes at the end of the main job (Figure~\ref{fig:kden_mjob_wage} and Figure~\ref{fig:kden_mjob_income}, respectively), are all quite similar. These results are partly attributable to the use of PMM for imputing the logarithm of positive monetary amounts. As discussed in Section~\ref{sec:MI_imp_eq}, this method does not rely on normality assumptions and effectively mitigates the occurrence of outliers in the imputed data.

\begin{figure}[htp]
\centering
\includegraphics[width=1.0\linewidth,height=0.40\textheight]{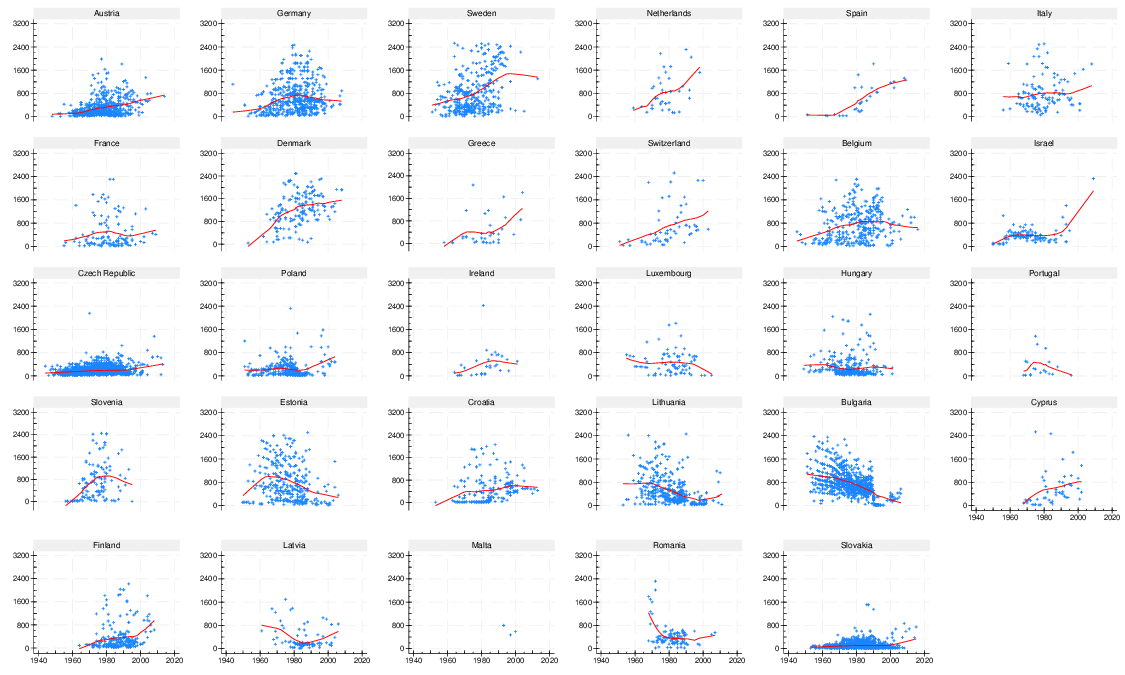}
\caption{Scatterplot between monthly maternity benefits and year of childbirth}
\label{fig:scatter-nch-mb}
\end{figure}

Figures~\ref{fig:kden_mb}  and~\ref{fig:kden_jsp_income} show that larger deviations between the observed
and imputed densities occur for monthly maternity benefits and first monthly incomes from self-employment.
Not surprisingly, these are also the variables with the lowest response rates (34 percent and 37 percent, respectively),
suggesting that the underlying nonresponse selection bias may represent a substantial source of nonsampling error.
For monthly maternity benefits, the imputed values assign relatively more probability mass to the right tail,
particularly between 500 and 1,500 Euros.
This pattern is consistent with the higher response rates in countries such as the Czech Republic and Slovakia,
where maternity benefits are generally much smaller than in other countries (see, e.g., Table~\ref{tab:des-mbn} and Figure~\ref{fig:scatter-nch-mb}).
These two countries alone account for about 40 percent of the observed sample, with average monthly maternity benefits of 100 and 175 Euros, respectively.
In contrast, the average monthly maternity benefit across the other countries is about 580 Euros.
Hence, it is not surprising that the distribution of imputed values places relatively greater weight on the right tail.

\begin{figure}[htp]
\centering
\includegraphics[width=1.0\linewidth,height=0.40\textheight]{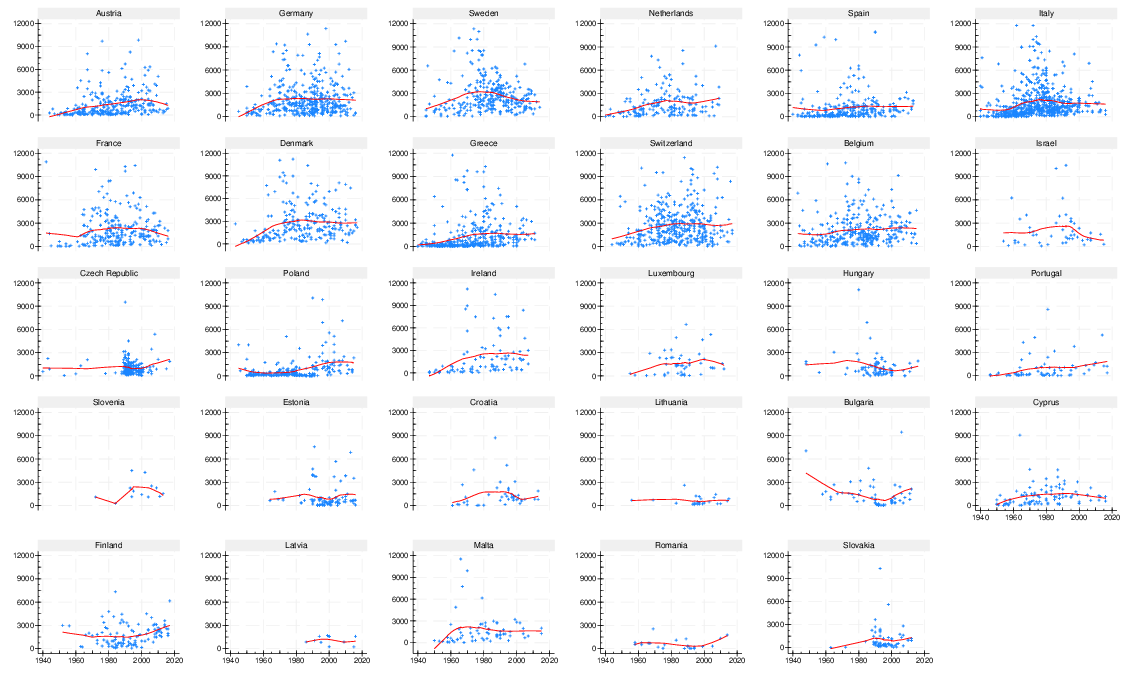}
\caption{Scatterplot of first monthly incomes from self-employment over time}
\label{fig:scatter-fmi}
\end{figure}

First monthly incomes from self-employment display the opposite pattern, with the imputed values assigning
relatively more probability mass to the left tail, particularly for amounts up to 1,000 Euros.
In this case, response rates are especially low in countries such as Slovenia, Israel, Poland, and Spain,
where the average first monthly income from self-employment tends to be smaller than in other countries
(see, e.g., Table~\ref{tab:des-fmi} and Figure~\ref{fig:scatter-fmi}).
Poland and Spain also report a relatively large number of self-employment spells; consequently,
the high number of imputed values from these two countries (around 1,000 cases each) substantially
influences the completed distribution of first monthly incomes.

\begin{figure}[htp]
\centering
\includegraphics[width=1.0\linewidth,height=0.40\textheight]{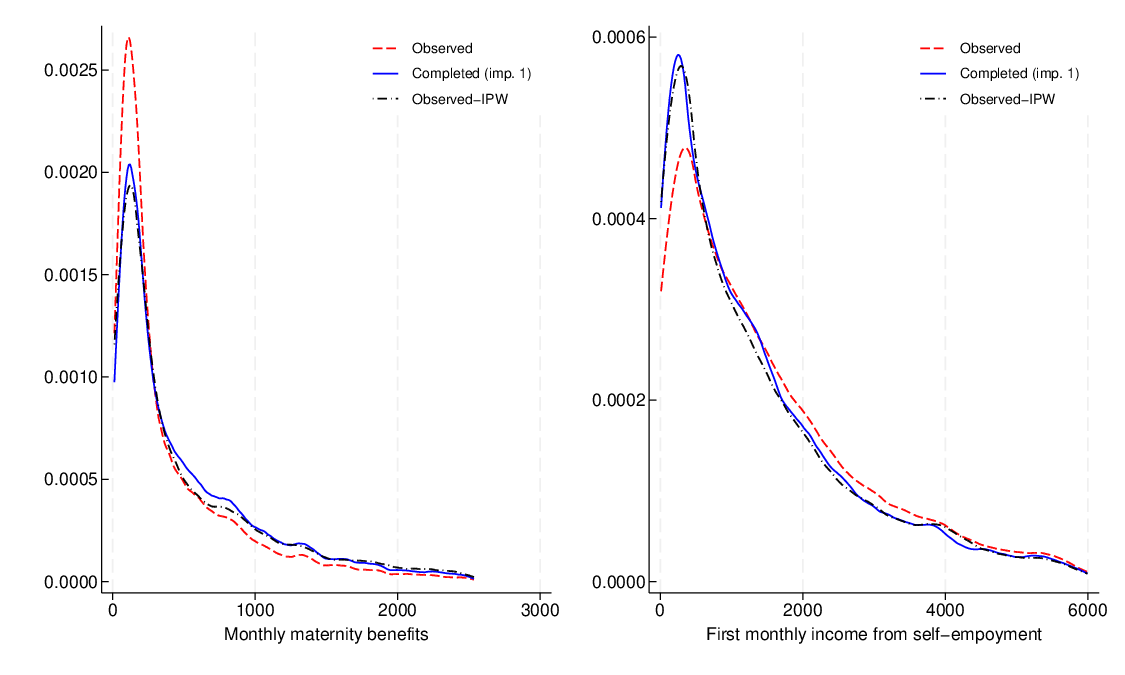}
\caption{Kernel densities of maternity benefits and monthly incomes from self-employment: observed sample,
completed samples, and observed sample with IPW}
\label{fig:kden2_mb_income}
\end{figure}

To gain additional insights into the determinants of the missing data mechanism and the adjustments introduced by multiple imputations, we now compare our approach with an inverse propensity score weighting (IPW) method. IPW is another widely used statistical technique designed to correct for nonresponse bias by weighting the observed sample with the inverse of the estimated response probabilities.
For this alternative approach, we first obtain the ML estimates of logit models for the response probability of each monetary outcome.
As with multiple imputations, these logit models are estimated separately by macro-region, using the set of predictors discussed in Section~\ref{sec:MI_predictors} and Appendix~A14.
The ML estimates reported in Appendix~C confirm the relevance of unobserved cross-country heterogeneity and provide further support for the comparisons between the observed and imputed data.
Provided that the response model is correctly specified, IPW and multiple imputations are expected to yield similar results.
Figure~\ref{fig:kden2_mb_income} presents the kernel density plots of monthly maternity benefits (right panel) and monthly incomes from self-employment (left panel).
Each panel focuses on the first imputation to compare the observed, completed (observed plus imputed), and IPW-weighted distributions.
As expected, the nonresponse bias adjustments based on the estimated propensity scores are very similar to those obtained under multiple imputation.
Appendix~C further shows that, for all other monetary variables, the results obtained from these three approaches remain closely aligned.

Next, we briefly turn to the issue of external validity.
As mentioned above, the main challenge in assessing external validity lies in ensuring the comparability of the outcomes of interest across different studies.
Differences in target population, country coverage, reference periods, question wording, and interview instruments can all substantially influence the results and limit the scope for direct comparison.
To minimize these concerns, we focus on a simple comparison of two monetary variables drawn from the regular waves of SHARE and from SHARELIFE.
Specifically, we restrict our attention to monthly wages from current employment and monthly pension benefits, both
expressed in euros at 2017 and adjusted for PPP to facilitate cross-country comparability.
%
\begin{table}[htp]
\caption{Distribution of current monthly wages and monthly pension benefits from regular SHARE waves and SHARELIFE}
\begin{center}
\begin{tabular}{lllrrrr}
\hline
\multicolumn{1}{l}{Variable} &
\multicolumn{1}{l}{Survey} &
\multicolumn{1}{l}{Wave} &
\multicolumn{1}{c}{Mean}  &
\multicolumn{1}{c}{p25}  &
\multicolumn{1}{c}{p50}  &
\multicolumn{1}{c}{p75}  \\
\hline
Current & SHARE & 1         &   2,127 &    887 &   1,848 &   2,902 \\
monthly  &      & 2         &   1,376 &    462 &   1,244 &   1,981 \\
wage &      & 4             &   1,343 &    474 &   1,100 &   1,946 \\
&      & 5                  &   1,509 &    604 &   1,346 &   2,110 \\
&      & 6                  &   1,416 &    540 &   1,244 &   1,991 \\
&      & 7                  &   1,306 &    333 &   1,098 &   1,914 \\
&      & 8                  &   1,264 &    336 &   1,015 &   1,832 \\
&      & 9                  &   1,218 &    392 &   1,021 &   1,764 \\
\cline{2-7}
&SHARELIFE & 3-7            &   1,653 &    938 &   1,463 &   2,156 \\
\hline
Monthly  & SHARE & 1        &   1,874 &    680 &   1,089 &   1,665 \\
pension & & 2               &   1,127 &    571 &     853 &   1,323 \\
benefits & & 4              &   1,179 &    512 &     803 &   1,290 \\
&  & 5                      &   1,155 &    585 &     882 &   1,377 \\
&  & 6                      &   1,067 &    520 &     805 &   1,247 \\
&  & 7                      &   1,120 &    633 &     915 &   1,322 \\
&  & 8                      &   1,037 &    521 &     773 &   1,186 \\
&  & 9                      &     959 &    481 &     755 &   1,155 \\
\cline{2-7}
& SHARELIFE & 3-7           &   1,156 &    506 &     946 &   1,597 \\
\hline
\end{tabular}
\end{center}
\label{tab:val-pb}
\end{table}

Table~\ref{tab:val-pb} presents summary statistics, including the mean, median, and 25th and 75th percentiles,
of the imputed distributions for these two monetary variables across all regular waves of SHARE and SHARELIFE.
For current monthly wages, SHARELIFE exhibits slightly higher values than those observed in the regular SHARE waves.
The only exception is Wave 1, in which all monetary variables were collected before taxes and social security contributions,
whereas in subsequent waves they were reported net of taxes and social contributions.
It is important to emphasize that the observed differences in these summary statistics are not necessarily attributable
to the imputation of missing values.
For instance, the country coverage of the SHARELIFE sample differs considerably from that of the regular SHARE Waves 2–6.
Moreover, differences in reference periods are likely to influence the comparison with the most recent waves, which
may reflect the negative economic effects of the COVID-19 pandemic in early 2020.

For monthly pension benefits, SHARELIFE again occupies an intermediate position (median about EUR 946; mean about EUR 1,156),
lying between the higher values recorded in SHARE Wave 1 (mean/median about EUR 1,874/1,089) and the lower values observed
in later waves (e.g., Wave 9 mean/median about EUR 959/755).


Dispersion is broadly comparable: for wages, the SHARELIFE interquartile range (IQR about EUR 1{,}218) is narrower than early SHARE (Wave 1 IQR $\approx$ EUR 2{,}015) and similar to later waves; for pensions, SHARELIFE's IQR (about EUR 1{,}091) is wider than the later SHARE waves but close to Wave 1.
In both surveys, means exceed medians, confirming right-skewed distributions.
Since amounts are standardized to a common year/currency, these gaps should be read primarily as differences in composition (countries, cohorts, and reporting frame—contemporaneous in SHARE vs. retrospective in SHARELIFE) rather than pure price-level effects.


\section{Conclusions}
\label{sec:CONCLUSIONS}

In this report, we have described the current state of the SHARELIFE-MI project,
which was developed to generate multiple imputations for missing values in the life-course data collected in SHARELIFE Waves 3 and 7.
We have presented the main design choices, the structure and implementation of the imputation model, and the diagnostic analyses conducted to assess its internal and external validity.
The resulting SHARELIFE-MI database provides a harmonized and fully documented resource that enables researchers to perform analyses accounting for item nonresponse while preserving the substantive and methodological features of the SHARELIFE study.
Future developments of the project will focus on extending the imputation model to account
explicitly for cross-country differences in the national sampling design.


\bibliographystyle{agsm}
\bibliography{references.bib}

\newpage

\section*{Appendix A. Building blocks of the imputation model}

This appendix provides detailed information on the specification and estimation of the 14 building blocks that constitute the SHARELIFE imputation model.
Each block comprises one or more groups of variables imputed through sequential applications of the FCS method.
Particular attention is devoted to the two-fold FCS algorithm developed for the imputation of monetary variables.

\subsection*{Appendix A1. Socio-demographics, proxy, and interviewer observations}

In the first block, we jointly impute 17 variables: the respondent’s age at the time of the interview, ISCED-97 educational coding, self-perceived health status, 11 binary indicators for proxy reporting across various SHARELIFE modules, and interviewer observations on the interview process (including whether third persons were present, whether the respondent requested clarifications, and whether the respondent understood the questions).
Overall, 2,471 observations (about 3\%) are incomplete and
58\% of the missing values are concentrated in ISCED-97 educational coding.
We use a Poisson model for the respondent’s age, a logit model for the presence of third persons, and an ordered logit model for all other outcomes.
The model is estimated separately by macro-region, with country dummies, household size, and binary indicators for being female, living with a spouse or partner, participating in the second SHARELIFE interview  included as exogenous and fully observed predictors.

\subsection*{Appendix A2. Family composition}

In the second block, we jointly impute four variables related to family composition: the number of natural children, adopted children, (cohabitant) spouses/partners, and non-cohabitant partners.
Overall, there are only 798 missing values, representing less than 1\% of the available cases.
Nevertheless, these missing values are crucial, as they determine complete sequences of missing data in the child and partnership histories.

We estimate a Poisson model for each of the four outcomes separately by macro-region.
In addition to country dummies, the set of predictors includes a second-order polynomial in the respondent's age at the time of the interview, household size, and binary indicators for being female, currently living with a spouse or partner, having a high educational level, reporting good health, participating in the second SHARELIFE interview, being interviewed with the help of a proxy respondent (either RC or RP modules), and being perceived by the interviewer as having a good understanding of the questions.
Missing values in these exogenous predictors are filled in with the imputed values obtained in the first block.

\subsection*{Appendix A3. Childhood circumstances}

In the third block, we jointly impute 11 variables describing childhood circumstances at age ten:
the number of rooms in the accommodation,
the number of people living in the household,
the number of books available at home,
binary indicators for the presence of specific household features (fixed bath, cold running water, hot running water, inside toilet, and central heating),
a binary indicator for school attendance,
and respondents’ relative standing in language and mathematics.
Missing values in these variables amount to 5,140 observations, corresponding to nearly 6\% of the available cases.
The variables with the highest rates of missingness are the number of rooms in the accommodation and the number of people living in the household, which account for 70\% and 61\% of all missing values, respectively.
We estimate a Poisson model for the number of rooms and the number of household members, an ordered logit model for the number of books and the respondents’ relative standing in language and mathematics, and a logit model for each binary indicator (household features and school attendance).
Respondents’ relative positions in language and mathematics are imputed through conditional imputations, since these variables are only defined for respondents who reported attending school.
Conditional imputations ensure that this characteristic of the observed data is preserved in the completed dataset, including both observed and imputed values.
The model for childhood variables is estimated separately by macro-region, using the same set of predictors as in the second block.

\subsection*{Appendix A4. Child history}

In the fourth block, we impute missing values in the sequence of child-history variables.
We focus on the demographic characteristics of natural children (gender, year of birth, survival status, and, where applicable, year of death), as well as information on maternity leaves (type of job interruption and duration of maternity leave), which is collected only from female respondents.
For adopted children, we impute analogous demographic characteristics: gender, year of adoption, survival status, and, where applicable, year of death.
The sequence of child-history variables includes up to 16 possible spells for natural children and 8 possible spells for adopted children.
Consequently, this block involves the imputation of a total of $(4+2)\times16 + 4\times8 = 128$ variables.
At this stage, we exclude the imputed values of maternity benefits, which are discussed separately in Appendix A13.1.
Missing values in the selected child-history variables amount to 3,866 observations, representing roughly 4\% of the available cases, with approximately 44\% of them concentrated in the duration of maternity leaves for natural children.

We proceed as follows.
First, we construct a chain for the gender and year of birth of natural children. Gender is modeled using a logit regression, while the year of birth is modeled using an interval regression for the respondent’s age at the time of the child’s birth.
%
This formulation of the latter variable is convenient, as it allows us to impose logical restrictions on its missing values.
Specifically, we assume that the respondent’s age at the time of the child’s birth is constrained to lie between 14 years and the minimum of 65 years and the respondent’s age at the time of the interview.
Hence, individuals cannot have a child outside the age range $[14, 65]$ years, and the age at childbirth is necessarily lower than the respondent’s age at the year of the interview.

The model for the first 5 natural children is estimated separately by macro-region, whereas data for later child spells are pooled across macro-regions.
This choice is justified by the fact that approximately 98\% of respondents report no more than five natural children.
Time dependence across children is modeled using an AR(1) structure,
in which each characteristic of child $c-1$ is allowed to influence the corresponding characteristic of child $c$ $(c=2,\ldots, 16)$.
The set of exogenous predictors corresponds to the complete set of variables detailed in Section~\ref{sec:MI_predictors}.
For higher-order children, a reduced subset of predictors is employed only when the available sample size becomes particularly small.

Next, we develop a chain to jointly impute the child’s survival status and, where applicable, the respondent’s age at the time of the child’s death.
We use a logit regression for the former variable and an interval regression for the latter, which is imputed conditionally on the child’s survival status.
Missing values for the respondent’s age at the time of the child’s death are constrained to lie between the respondent’s age at childbirth and the age at the time of the interview.
As with the other child characteristics, we estimate the model for the first 5 natural children separately by macro-region and pool data from multiple macro-regions for the subsequent child spells.

Similar considerations apply to the imputation model for maternity leaves, which consists of a multinomial logit model for the type of job interruption and an interval regression for the duration of maternity leave.
Both variables are asked only of female respondents, and the latter is defined only for those who reported a temporary job interruption in the former question.
The model for the first 5 child spells is estimated separately by macro-region, including country dummies and other child characteristics (gender and year of birth).
For subsequent child spells, data from multiple macro-regions are pooled, and the number of predictors is reduced when necessary to address issues of small sample size and convergence problems.

Finally, we construct a chain for the gender and year of adoption of adopted children, and a separate chain for their survival status and, where applicable, the respondent’s age at the time of the child’s death.
Compared with natural children, we simplify the specification of the imputation model to account for the limited sample size of adopted children.

\subsection*{Appendix A5. Partner history}

The fifth block covers the partner history variables collected in the RP module. Particularly, this block reconstructs, for up to eight cohabitant partners or spouses, the age at which cohabitation began, whether the relationship ended, the age at which it ended, and whether the couple married. Ages are expressed on a common scale: whenever the questionnaire provides calendar years, we first convert them to ages as reported year minus birth year and then treat the resulting ages as the imputation targets wherever timing is missing or unusable. All imputations use FCS imputation, are estimated separately by macro-region with country fixed effects, and condition on the standard socio-demographic predictors together with within-block auxiliaries.

The chain follows the episode structure implied by the questionnaire. For each partner $j=1,\ldots,8$, the start of cohabitation is modeled on its natural support with interval regression. For partner $j$ the start age of cohabitation is imputed by interval regression with the lower bound set to the end age of partner $j-1$ (when available), thereby enforcing non-overlap across consecutive relationships. For the first partner the lower bound is the generic minimum (e.g. 14 years old), while for later partners any missing previous end age is handled with the fallback in the code before applying the upper bound at the interview age.

Marriage is modeled with a logit and is conditioned on the timing information so that the probability of marriage coheres with the observed or imputed spell. Temporal logic is imposed to avoid overlaps and to preserve sequence across partners. Within each partner, the end age cannot precede the start age. Across partners, later relationships cannot begin before earlier ones have ended; when both ages are missing, weak ordering is induced by including adjacent spells as predictors and, where necessary, by tightening the interval bounds so that the imputed start of partner $j+1$ is not earlier than the imputed or observed end of partner $j$. This yields non-overlapping, chronologically ordered spells of cohabitation and termination. Because marriage can occur within an ongoing relationship, we do not force marriage to coincide with the start or end of cohabitation; instead, marriage status is allowed to depend on the cohabitation timing and the respondent's characteristics, and consistency checks after imputation ensure there is no marriage recorded outside a plausible cohabitation window.

All equations draw on the common covariate set (age and age squared, sex, education, household size, self-reported health, wave indicators, proxy/interviewer flags) and on within-block auxiliaries such as the presence and timing of adjacent partner spells. Estimation proceeds by macro-region with country fixed effects; when specific partner-order cells become too sparse to sustain the full specification, we pool across macro-regions or reduce the predictor set to the core controls while retaining the same constraints on support and ordering. Post-imputation diagnostics verify that ages lie in range, that the ordering across partners is respected, and that end-before-start or overlapping spells do not arise.

\subsection*{Appendix A6. Accommodation history}

The sixth block covers accommodation histories recorded in the AC module. We begin by reconstructing the spell sequence under the maintained assumption that every respondent experienced at least two accommodation spells: a baseline spell associated with the birth residence and a second spell corresponding to the first move after birth. Timing is expressed in age, obtained by converting any reported calendar year into age as the difference between the reported year and the birth year; these ages are the targets of imputation when missing or unusable. Spells are contiguous by construction: the start of a new residence coincides with the end of the previous one, so there are no gaps or overlaps. The first spell anchors the history; the start age of spell $t+1$ is constrained to equal the end age of spell $t$. This ``end-equals-next-start'' rule is enforced through the bounds used in the timing models. We also classify spells as short-term when their duration is at most six months; by definition, short-term spells begin and end within the same calendar year and therefore share the same age in our timing scale, and the next spell inherits that age as its start, preserving contiguity. An end-of-history indicator marks the last spell in a respondent's sequence. Once this flag is reached - based on the reported or imputed number of spells - all higher-order spells (e.g., $k+1,k+2,\ldots$ when the last realized spell is $k$) are treated as structurally missing and are excluded from imputation.

Within this framework, start and end ages are imputed by interval regression on feasible supports explained above. Particularly, imputation strategy proceeds by imputing each episode in a separate chain which depends only on the previous one. In total there are 31 episodes to be imputed. Moreover, for each episode the indicator whether the accommodation episode is transitory or not is imputed by logit model.

After timing of accommodation has been imputed, spell attributes - such as private versus non-private residence, type of residence, country/region/area, and whether property was purchased or sold during the spell - are imputed conditional on spell existence and timing, allowing attributes to borrow information from adjacent spells. All imputations follow the FCS approach, are estimated separately by macro-region with country fixed effects, and condition on the common socio-demographic predictors. When late  spells are too sparse to support the full specification (e.g only 5\% of the sample has more than 10 accommodation episodes)  we fall back to pooled or lean models while preserving the same contiguity and non-overlap constraints expressed in ages.

\subsection*{Appendix A7. Work history}

The seventh block includes approximately 150 work-history variables collected in the RE module, spanning full-time education and various aspects of the respondent’s career.
By design, the RE module records information for up to 20 job spells.
It partially overlaps with the WQ and DQ modules, which collect information on work quality history and disability benefits.
FCS imputations for variables within these two modules are discussed in Appendices~A8 and~A9.
Likewise, the imputation of monetary variables is deferred to Appendix~A14.

Missing values in the selected work-history variables amount to 15,088 observations, representing roughly 16\% of all available cases.
About 82\% of these missing values are concentrated in variables related to full-time education, the starting years of job spells, and the job situation at the end of each spell.
Most missing values tend to occur in the first few job spells, which are also the oldest ones.
This pattern is not entirely surprising, as in addition to `don’t know' and `refusal' responses, we also impute possible inconsistencies between the reported end date of full-time education and the beginning of work histories.

We begin by sequentially imputing a preliminary set of variables that determine eligibility conditions for the sequence of job spells.
This requires estimating the following models:
a logit model for school attendance;
a multinomial logit model for the job situation at age 15 among respondents who did not attend school;
a logit model for having ever engaged in paid work;
an ordered logit model for the start of the first paid job;
a multinomial logit model for the job situation after full-time education;
an interval regression for the number of recorded job spells;
a multinomial model for the exit reason from the sequence of job spells;
and an interval regression for the job spell corresponding to the main job in the respondent’s career.
Most of these models are estimated separately by macro-region, including both country fixed-effects and the full set of predictors described in Section~\ref{sec:MI_predictors}.
The only exception is the multinomial logit model for the job situation at age 15, which is estimated for the 1,523 respondents who did not attend school. For this variable, we pool data across macro-regions and replace country dummies with macro-region dummies.

Next, we consider the age at which the respondent completed full-time education, as well as the starting and ending years of the various job spells.
As usual, to impose logical constraints on this sequence of variables, we model the starting and ending years of the job spells using interval regressions based on the respondent's age at the time each spell occurred.
Within these models, information from past and future job spells is used to define the range of admissible values for the imputation of the $h$th spell in the sequence ($h = 1, 2, \ldots$).
This ensures that job spells start after the completion of full-time education and occur in sequential order.
We estimate the models for the first seven job spells separately by macro-region, whereas data from multiple macro-regions are pooled for the subsequent job spells.
To avoid multicollinearity, we exclude the dummy variable for high educational level from the model for the age at which the respondent completed full-time education.

In the same fashion, we then consider the full sequence of 20 history variables for the job situation at the end of each job spell, as well as possible time gaps between different spells.
The former are estimated using multinomial logit models, while the latter are estimated using ordered logit models.
This block of the SHARELIFE imputation model concludes with a chain of multinomial logit models for the joint modeling of industry, type of job (employed, civil servant, or self-employed), and part-time versus full-time work.
Most of these models are again estimated separately by macro-region (at least for the first six to seven job spells), using an AR(1) structure in which features of previous job spells are included as predictors of the current job spell.

\subsection*{Appendix A8. Work quality history}
The eighth block covers the work-quality variables collected in the WQ module. The module asks about two job references: the current job and the main job over the career. For each reference, respondents answer 12 ordinal (Likert-type) items on working conditions: work is physically demanding, work environment is uncomfortable, work has heavy time pressure, work is emotionally demanding, work involves conflicts, work allows little freedom to decide, work allows development of skills, work gives recognition, salary is adequate, support is adequate, work atmosphere is good, and health risks in the workplace are reduced.
As described in the methodological section, we use a Fully Conditional Specification approach: each variable is imputed from a univariate model conditional on the others, iterating the chain and repeating to obtain $M$ independent imputations. This means that the 12 items for each working conditions are jointly imputed. Given the ordinal nature of the outcomes, each WQ item is imputed using an ordered logit model. Imputations are run separately by macro-region and include as predictors age (and age squared), sex, education, self-reported health, household size, wave indicators, proxy/interviewer flags, and country fixed effects, alongside the other WQ indicators for the same job reference; imputations are further restricted to respondents eligible under the module's skip patterns.

\subsection*{Appendix A9. Disability benefits}

The ninth block covers the Disability benefits module. In particular it focuses on the imputation of i) three labor-market conditions linked to  disability (ever leaving a job because of disability, taking a temporary leave of absence, and limiting working hours), and ii) four insurance/pension outcomes that trace the public and private disability protection pathways (ever applying for a public disability pension and ever receiving it, ever purchasing private disability insurance, ever applying for a private disability insurance claim, and ever receiving a private disability insurance payout). All outcomes are binary and are imputed within the Fully Conditional Specification framework, estimated separately by macro-region with country fixed effects and conditioned on the standard socio-demographic predictors, plus cross-block auxiliaries from application to private disability and insurance

The imputation specification proceeds in blocks that mirror the conceptual ordering of the processes. A first group imputes the three work-adjustment indicators by logit, allowing them to inform one another while conditioning on age and age squared, sex, education, household size, self-reported health, wave, and proxy/interviewer flags, plus cross-block auxiliaries from the health-history module (e.g., serious injury and illness-burden measures) and the work-quality module. This preserves the idea that exits, temporary leaves, and reduced hours are related yet distinct adjustments whose incidence varies with health and job context.

A second group focuses on the public pension pathway. Application for a public disability pension is imputed by logit on the same predictors base, augmented by the work-disability indicators so that the likelihood of applying can reflect prior labor-market disruptions. Receipt of a public disability pension is then imputed by logit under a strict routing rule: receipt is structurally missing for respondents who did not apply, and is modeled only among applicants. This enforces the administrative sequence implied by the questionnaire and prevents infeasible combinations.

A third group addresses the private insurance pathway. Purchase of private disability insurance is imputed by logit using the common predictors and the work-disability indicators. Application for a private claim and receipt of a private payout are then imputed as conditional logit models that respect the private sequence: application is modeled only among those who have purchased coverage, and receipt is modeled only among those who have applied. These structural restrictions are implemented as routing conditions so that non-eligible cases remain absent and never enter the chain. As in the public pathway, the equations exploit information from earlier blocks—particularly health and work-quality—to stabilize prediction in sparsely populated strata. These three blocks are jointly imputed separately by macro-region.

\subsection*{Appendix A10. Financial history}
The tenth block covers the financial history variables collected in the FS module. This block reconstructs lifetime exposure to four financial instruments—stocks or shares, mutual funds, life insurance, and business ownership—and the timing of first uptake for each. For every instrument we observe an ever/never question (e.g., “ever had any stocks or shares”) and, when available, a calendar year of first uptake. To place timing on a common scale, reported years are converted to ages by subtracting the respondent’s birth year; these age-at-first-uptake variables are the targets of imputation wherever the year is missing or unusable. Ages are relevant—and therefore imputed—only when the corresponding ever/never indicator is positive; otherwise they remain structurally missing by design.

Imputation follows the FCS framework used elsewhere. We iterate a chain in which the four ever/never indicators are modeled with logit links, and the age-at-first-uptake variables are modeled with interval regression, conditioning each equation on the other financial variables in the block (including the other age-at-first-uptake variables) and the standard predictors. Interval bounds impose a sensible lower limit at 18 years (legal adulthood in most countries) and an upper limit at the interview age.

To respect heterogeneity in financial markets and reporting across Europe, imputations are run separately by macro-region with country fixed effects. The predictor set mirrors our common specification—age and age squared, sex, education, household size, self-reported health, wave indicators, and proxy/interviewer flags—augmented by the other financial variables in this block so that, for example, information on mutual fund participation can inform the imputation of stock ownership, and vice versa.

\subsection*{Appendix A11. Health history}	

The eleventh block covers the health history variables collected in the HS module. The block brings together three strands of information: (i) childhood health-self-assessed health in childhood and whether the respondent was confined to bed/home or hospitalized for one month or longer; (ii) serious injury - whether the respondent ever experienced a physical injury leading to disability and the age at injury; and (iii) periods of ill health over the life course - the number of periods and, for up to three periods, the start age, whether the period ended, the end age (if applicable), and counts of illnesses recorded in two pre-specified sets.

Imputation follows the Fully Conditional Specification framework used elsewhere: each outcome is modeled conditionally on the others in this block and on the common predictors, iterating the chain and repeating to obtain $M$ completed databases. Estimation is carried out separately by macro-region with country fixed effects. Particularly, childhood self-assessed health is treated as ordinal and imputed via ordered logit. The indicators for month-long confinement and hospitalization are binary and imputed via logit. Predictors include standard demographics (age and age squared, sex, education, household size), self-reported health at interview, wave indicators, proxy/interviewer flags and country effects.

Regarding serious injury, the ever/never indicator for a physical injury leading to disability is imputed with a logit model. When a calendar year of injury is reported, we first derive age at injury as (injury year - birth year); the age variable is then the imputation target wherever this is missing or unusable. Age at injury is imputed only for respondents with an injury; otherwise it remains structurally missing. Imputation uses interval regression with plausibility bounds: a lower bound at a sensible minimum (e.g., 0 for conditions reported since birth, or a conservative mid-adolescence threshold otherwise) and an upper bound at the interview age.

We model the two summary counts of childhood illnesses as Poisson outcomes (support 0–7). Predictors include childhood self-assessed health, indicators for being confined to bed/home for more than one month during childhood, and the number of period-specific illness.

Finally, the number of periods is treated as an ordered categorical outcome with six levels - None, One, Two, Three, More than three, and Been ill/with disabilities for all/most of life - summarizing the lifetime burden of ill-health. We model this variable with an ordered logit, reflecting its natural ordering. For each period $j=1,2,3$, we also impute period-specific details: the start age, whether the period ended, the end age (if it ended), and the number of illnesses recorded in the two predefined sets. Ages are obtained by first converting reported years into ages (reported year - birth year) and then imputing missing values via interval regression with plausibility bounds (e.g. if the period of illness ended, then end age cannot be larger than start age). The period-specific illness counts are modeled using Poisson regression, counting up to seven diseases from the predefined lists respondents can choose from. Skip patterns are respected throughout: period-specific variables are defined only for periods implied by the value of the ordered count.

Imputations are run by macro-region. All equations draw on the common predictor set used throughout the appendix - age and age squared, sex, education, household size, self-reported health, wave indicators, proxy/interviewer flags—plus country fixed effects and the other HS variables in this block (including lagged/adjacent period information where applicable). ; when cells are sparse, we follow the project’s pooling rules.

\subsection*{Appendix A12. Health care}

The twelfth block covers lifetime utilization of preventive and routine care in three areas: childhood vaccinations, regular dental care, and regular blood-pressure checks. For each domain, the imputed variables cover (i) high-level participation (ever/regular, always-regular), (ii) timing (age at start) and/or frequency of checks, 
Missingness arises from routing (reasons asked only if no uptake/regularity), partial nonresponse within batteries, and absent timing information.

Childhood vaccination is a binary indicator imputed with a logit model. The nine reasons for no vaccination (affordability, insurance coverage/ownership, time constraints, information, social norms, access, perceived necessity, other) are also binary items. Within the FCS framework, vaccination status and the reason items are jointly imputed. Reasons are imputed only for respondents who report not being vaccinated in childhood and are treated as structurally missing otherwise.

For regular dental care three aspects are imputed: when regular checkups began, how frequently they occur, and whether checkups were always regular. The age at start is modeled only for respondents who report having regular checkups. When a calendar year is available, we first convert it to age (reported year minus birth year) and then impute missing values using interval regression with plausible bounds (non-negative and not exceeding the interview age). Frequency of checkups is an ordered outcome (from ``never/rarely'' up to ``very often'') and is imputed with an ordered-logit specification, conditional on regular/always-regular status and standard predictors. The ``always regular'' indicator is binary and is modeled with a logit, while the frequency regular dentist is modeled via a ordered logit given the ordinal nature of the responses. All these variables are  modeled by macro-region.


Finally, for the variables on regular blood-pressure checks, imputation follows the same logic as for dental care and is estimated separately by macro-region, with country fixed effects.

\subsection*{Appendix A13. General life}
The thirteenth block covers the General Life module, and in particular the imputation of: periods of happiness, periods of stress, periods of financial hardship, periods of hunger, and experiences of discrimination and persecution, including downstream consequences (job disruption, difficulties finding a job) and episodes of dispossession with the type of property affected. For each period-type we impute a yes/no indicator, the age at start, an end indicator, and the age at end (when applicable). For persecution we also record the main reason, job-related consequences, the number of dispossession episodes (up to five), the age at each episode, and the property categories involved.

Whenever the questionnaire supplies a calendar year for an onset or an end, timing is first converted to age by subtracting birth year; these derived ages are the targets of imputation when missing or unusable. All imputations follow the FCS framework used throughout the appendix, are estimated separately by macro-region with country fixed effects, and condition on the common set of socio-demographic predictors (age and age squared, sex, education, household size, self-reported health, wave indicators, and proxy/interviewer flags) alongside block-specific auxiliaries.

In the first FCS block, we impute the four ever/never indicators for the core domains: period of happiness, period of stress, period of financial hardship, and period of hunger. Each is modeled with a logit, using the block-specific predictor sets and the common socio-demographic predictors, with estimation by macro-region and country fixed effects.

In the second FCS block we impute the timing for each domain that is present. For happiness, stress, financial hardship, and hunger, we impute the age at start using interval regression on a plausible support (with a lower bound at zero and an upper bound equals to end indicator or the interview age), the end indicator with a logit model, and - when the period is reported to end - the age at end via interval regression, requiring that the end age is not earlier than the start age. Because the questionnaire provides calendar years, we first convert these to ages (reported year minus birth year) and then impute any missing or unusable ages. All equations condition on the ever/never status from Block 1, the common socio-demographic predictors, and within-block auxiliaries, and are estimated by macro-region with country fixed effects.

In the next FCS block we address discrimination and persecution. The indicator of having been discriminated against or persecuted is treated as a binary outcome and imputed with a logit model. The stated main reason for persecution is nominal and is modeled with a multinomial specification; it is estimated jointly with the discrimination indicator so that the two reinforce each other and rare reason categories can borrow strength from the predictors and regional patterns. Labor-market consequences are handled conditionally on discrimination: the indicators for being forced to stop working and for experiencing difficulties in finding a job are both binary logit models that include the main-reason categories among their predictors. The timing of the first episode of job-search difficulty is imputed only for respondents who report such difficulties; when the questionnaire provides a calendar year we first convert it to age as the difference between the reported year and the birth year, and then we impute missing values using interval regression with bounds that require non-negative ages not exceeding the interview age. All equations condition on the common socio-demographic set. Estimation is conducted separately by macro-region with country fixed effects. Skip logic is enforced throughout: if discrimination is not reported, the reason and all job-consequence variables remain structurally missing and are excluded from the chain; if job-search difficulty is not reported, its timing is likewise left structurally missing.

The final FCS block is focused on the imputation of dispossession with the lifetime number of episodes and then, conditional on that count, reconstruct the age at which each episode occurred (up to five) together with the types of property affected at each episode. The total count is treated as an ordered categorical outcome and imputed with an ordered-logit specification on the usual predictor set (socio-demographics and macro-region with country fixed effects). This count governs eligibility downstream: episode-specific fields exist only when the count implies that the corresponding episode occurred; otherwise they remain structurally missing and are excluded from the chain.

For timing of each of episode, whenever the questionnaire provides a calendar year for an episode we first convert it to age by subtracting the birth year and then impute missing or unusable values by interval regression on a plausible support bounded below at eighteen and above at the interview age. Within each episode, the property categories (business/company, house/building, farmland/land, flat/apartment, money/assets) are binary check-boxes estimated as logits, allowing multiple categories to be selected for the same event. Imputation of timing variables is jointly modeled across episodes, but separately from property indicators. All equations are estimated separately by macro-region with country fixed effects, alongside the common socio-demographic controls. However, when sample sizes are thin for specific episodes (e.g. 4 and 5) or property-type cells, we use a graded back-off strategy rather than forcing the full specification. If a model cannot be stably estimated within a macro-region, we first pool across macro-regions and retain country effects only where estimable; if necessary, we replace country fixed effects with macro-region dummies, or we reduce the predictor set to the common socio-demographics. In the rare cases where even this lean specification is unstable, we switch to simple logit with minimal predictors or sampling from a uniform distribution. This ensures identification without introducing artifacts, while keeping imputations as close as possible to the intended design.

\subsection*{Appendix A14. Monetary variables}

Our last block of imputations addresses the most challenging issue, namely that of missing data on monetary variables.
As discussed in Section~\ref{sec:money_conv}, we do not model the information on the reference period, amount, and currency separately.
Instead, we directly model the logarithm of PPP-adjusted amounts of the following monetary variables:
\begin{itemize}
\item the sequence of monthly maternity benefits for natural children $Y_{1h}^{}$  ($h=1,\ldots,13$);
\item the sequence of first monthly wages $Y_{2h}^{}$  ($h=1,\ldots,20$);
\item the sequence of first monthly incomes from self-employment $Y_{3h}^{}$  ($h=1,\ldots,16$);
\item monthly pension benefits when retired $Y_4^{}$;
\item current monthly wages $Y_{5}^{}$;
\item current monthly incomes from self-employment $Y_{6}^{}$;
\item monthly wages at the end of the main job $Y_{7}^{}$; and
\item monthly incomes from self-employment at the end of the main job $Y_{8}^{}$.
\end{itemize}
By design, these monetary variables are collected under specific eligibility conditions.
For example, information on maternity benefits is gathered only from women who reported a temporary job interruption due to childbirth,
wages are defined only for employees and civil servants,
incomes from work are defined only for the self-employed,
and pension benefits are collected only from individuals who are retired from work.

Rather than modeling all these variables jointly through a standard FCS algorithm, we simplify the imputation process by iterating across four chains of variables:
one chain for the sequence of maternity benefits $Y_{1h}^{}$;
one chain for the sequence of first monthly wages $Y_{2h}^{}$;
one chain for the sequence of first monthly incomes $Y_{3h}^{}$;
and one chain for the remaining monetary variables $(Y_{4}^{}. \ldots, Y_{8}^{})$.
%
At iteration $t\ge 1$, the imputed values of $Y_{1h}^{}$ are obtained by drawing from a conditional density
of the form
\begin{equation}
Y_{1h}^{(t)} \sim f_1^{}\left(Y_{1h}^{}\big|\mathcal{H}_{1h}^{(t)}; \psi_1^{}\right),
\label{eq:chain_1}
\end{equation}
where $\psi_1^{}$ is a vector of parameters and the conditioning set
$$
\mathcal{H}_{1h}^{(t)}=\left\{\bar Y_{2}^{(t-1)}, \bar Y_{3}^{(t-1)}, Y_{4}^{(t-1)}, \left(Y_{5}^{(t-1)}+Y_{6}^{(t-1)}\right), \left(Y_{7}^{(t-1)}+Y_{8}^{(t-1)}\right), O_{1,h-1}^{(t)}, Y_{1,h-1}^{(t)}, Z_{1}^{}\right\},
$$
includes the averages of the $j$th monetary sequences
$\bar Y_{j}^{(t-1)}=H_j^{-1} \sum_{h=1}^{H_j^{}} Y_{jh}^{(t-1)}$,
the ownership indicator $O_{1,h-1}^{(t)} = 1(Y_{1,h-1}^{(t)} > 0)$ for the previous child spell, and the predictor set $Z_1^{}$ described in Section~\ref{sec:MI_predictors}, now augmented with a quadratic in the respondent’s age at childbirth.
This specification assumes that the sequence of maternity benefits $Y_{1h}^{}$ follows an AR(1) process, in which both the ownership status and the amount in the previous child spell may influence the amount in the current spell.\footnote{
\ The initial condition of the AR(1) process is assumed to be zero.}
To preserve the correlations with the other monetary variables, we also allow the maternity benefits to depend on the average of the first monthly wages, the average of the first monthly incomes, the pension benefits, the current monthly salary (wages plus income from work), and the salary at the end of the main job.
The imputation model for $Y_{1h}^{(t)}$ is specified as a two-part model that consists of a logit model to predict ownership (i.e., whether the outcome is zero or positive), followed by a predictive mean matching for the positive part of the distribution.
In this case, the models for the first three child spells are estimated separately by macro-region, while data from multiple macro-regions are pooled for the subsequent spells.
The number of nearest neighbors used in predictive mean matching is set to $10$ up to the seventh child spell, and is then gradually reduced as the sample size becomes smaller.

The conditional density for the sequence of first monthly wages is specified as
\begin{equation}
Y_{2h}^{(t)} \sim f_2^{}\left(Y_{2h}^{}\big|\mathcal{H}_{2h}^{(t)}; \psi_2^{}\right),
\label{eq:chain_2}
\end{equation}
with the conditioning set given by
$$
\mathcal{H}_{2h}^{(t)}=\left\{\bar Y_{1}^{(t)}, \bar Y_{3}^{(t-1)}, Y_{4}^{(t-1)}, Y_{5}^{(t-1)}, Y_{7}^{(t-1)}, O_{2,h-1}^{(t)}, Y_{2,h-1}^{(t)}, Z_2^{}\right\},
$$
where $O_{2,h-1}^{(t)} = 1(Y_{2,h-1}^{(t)} > 0)$ and $Z_2$ comprises the predictors described in Section~\ref{sec:MI_predictors}, now augmented with a quadratic polynomial in the respondent’s age at the beginning of the $h$th job spell.
This specification mirrors the model used for the sequence of maternity benefits.
The main difference compared with equation~\eqref{eq:chain_1} is that income from work, both in the current and main job, is assumed not to affect the first monthly wages.
Another key difference is that, unlike maternity benefits, we only need to model the PPP-converted amounts conditional on ownership. Any missing values for ownership (i.e., the choice between employment, self-employment, and civil service) were already imputed in the work-history block (see Appendix~A7). Imputed amounts are obtained using predictive mean matching, separately by macro-region for the first six spells and pooled across macro-regions for the subsequent spells.

Similarly, for the sequence of the first monthly incomes, we assume that
\begin{equation}
Y_{3h}^{(t)} \sim f_3^{}\left(Y_{3h}^{}\big|\mathcal{H}_{3h}^{(t)}; \psi_3^{}\right),
\label{eq:chain_3}
\end{equation}
with
$$
\mathcal{H}_{3h}^{(t)}=\left\{\bar Y_{1}^{(t)}, \bar Y_{2}^{(t-1)}, Y_{4}^{(t-1)}, Y_{6}^{(t-1)}, Y_{8}^{(t-1)}, O_{3,h-1}^{(t)}, Y_{3,h-1}^{(t)}, Z_3^{}\right\},
$$
$O_{3,h-1}^{(t)} = 1(Y_{3,h-1}^{(t)} > 0)$, and $Z_3^{}\equiv Z_2^{}$.
In this case, we assume that wages from both the current and main job do not affect the first monthly incomes.
As for the first monthly wages, we only need to model the PPP-converted amounts conditional on ownership.
Due to the limited sample size, these imputed values are obtained through predictive mean matching on data pooled across macro-regions.

Finally, we specify a chain for the last five monetary variables:
\begin{equation}
Y_{j}^{(t)} \sim f_j^{}\left(Y_{j}^{}\big|\mathcal{H}_{jh}^{(t)}; \psi_j^{}\right),
\qquad (j=4,\ldots,8),
\label{eq:chain_4}
\end{equation}
where
$\mathcal{H}_{4h}^{(t)}=\left\{\bar Y_{1}^{(t)}, \bar Y_{2}^{(t)}, \bar Y_{3}^{(t)}, \left(Y_{7}^{(t-1)}+Y_{8}^{(t-1)}\right),  Z_{4}^{}\right\}$,
$\mathcal{H}_{5h}^{(t)}=\mathcal{H}_{6h}^{(t)}=\left\{\bar Y_{1}^{(t)}, \bar Y_{2}^{(t)}, \bar Y_{3}^{(t)}, Z\right\}$,
and
$\mathcal{H}_{7h}^{(t)}=\mathcal{H}_{8h}^{(t)}=\left\{\bar Y_{1}^{(t)}, \bar Y_{2}^{(t)}, \bar Y_{3}^{(t)}, Y_{4}^{(t)}, Z_{7}^{}\right\}$.
The predictors in $Z$ correspond to the variables described in Section~\ref{sec:MI_predictors}; those in $Z_4$ additionally include a quadratic polynomial in the respondent’s age at the end of the last job spell, whereas those in $Z_7$ include a quadratic polynomial in the respondent’s age at the end of the main job.
The more parsimonious specifications of $\mathcal{H}_{5h}^{(t)}$ and $\mathcal{H}_{6h}^{(t)}$ reflect the fact that the reference period for $Y_5^{}$ and $Y_6^{}$ is the interview year, so a polynomial in the respondent’s age at the reference period would coincide with that already included in $Z$.
In addition, these two monetary variables are observed only for respondents who were employed or self-employed in the interview year.
The imputation model for $Y_{4}^{}$ is again specified as a two-part model, since it requires imputing missing values for both ownership and amounts, whereas for the other variables, only the missing amounts need to be imputed.
For the latter, predictive mean matching is consistently used to prevent the occurrence of outliers in the imputed values.
All models are estimated separately by macro-region.

Our two-fold FCS algorithm proceeds as follows:
\begin{enumerate}
\item Compute the FCS imputations of $Y_{1h}^{}$ $(h = 1, \ldots, 13)$ using model~\eqref{eq:chain_1};
\item Compute the FCS imputations of $Y_{2h}^{}$ $(h = 1, \ldots, 20)$ using model~\eqref{eq:chain_2};
\item Compute the FCS imputations of $Y_{3h}^{}$ $(h = 1, \ldots, 16)$ using model~\eqref{eq:chain_3};
\item Compute the FCS imputations of $(Y_{4}^{}, Y_{5}^{}, Y_{6}^{}, Y_{7}^{}, Y_{8}^{})$ using model~\eqref{eq:chain_4};
\item Iterate the previous four steps over a predefined burn-in period, and then retain the final set of imputed values for all monetary variables from the last iteration.
\end{enumerate}

\section*{Appendix B. Scatterplots of monetary variables}


\begin{figure}[htp]
\centering
\includegraphics[width=1.0\linewidth,height=0.40\textheight]{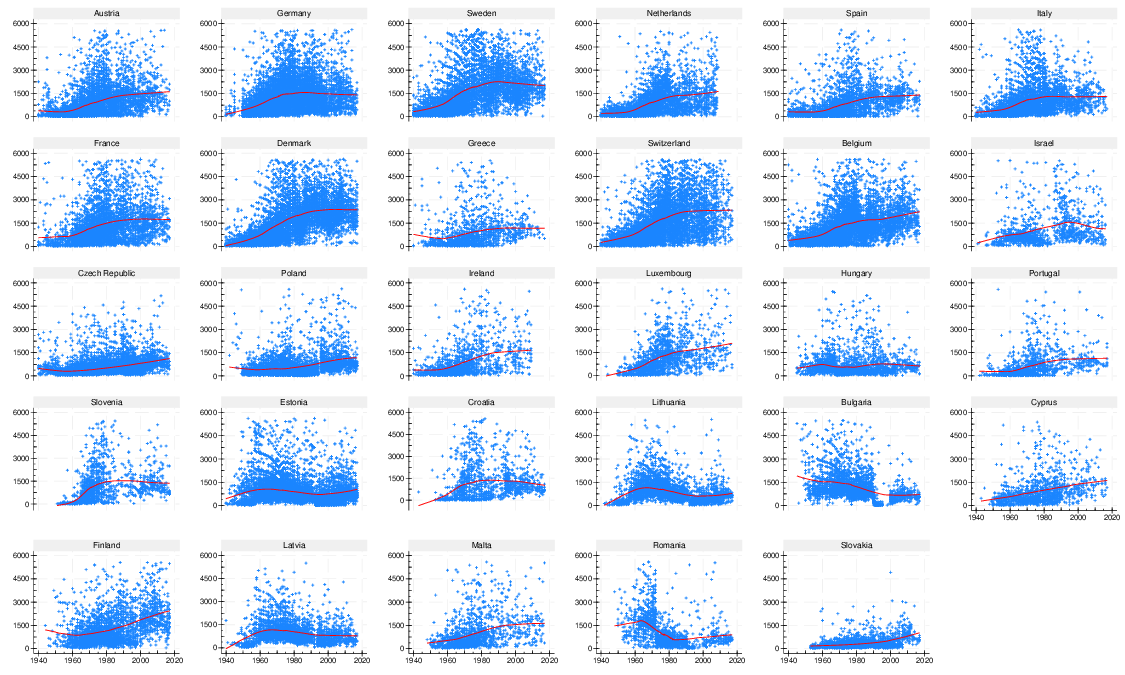}
\caption{Scatterplot of first monthly wages over time}
\label{fig:scatter-fmw}
\end{figure}

\begin{figure}[htp]
\centering
\includegraphics[width=1.0\linewidth,height=0.40\textheight]{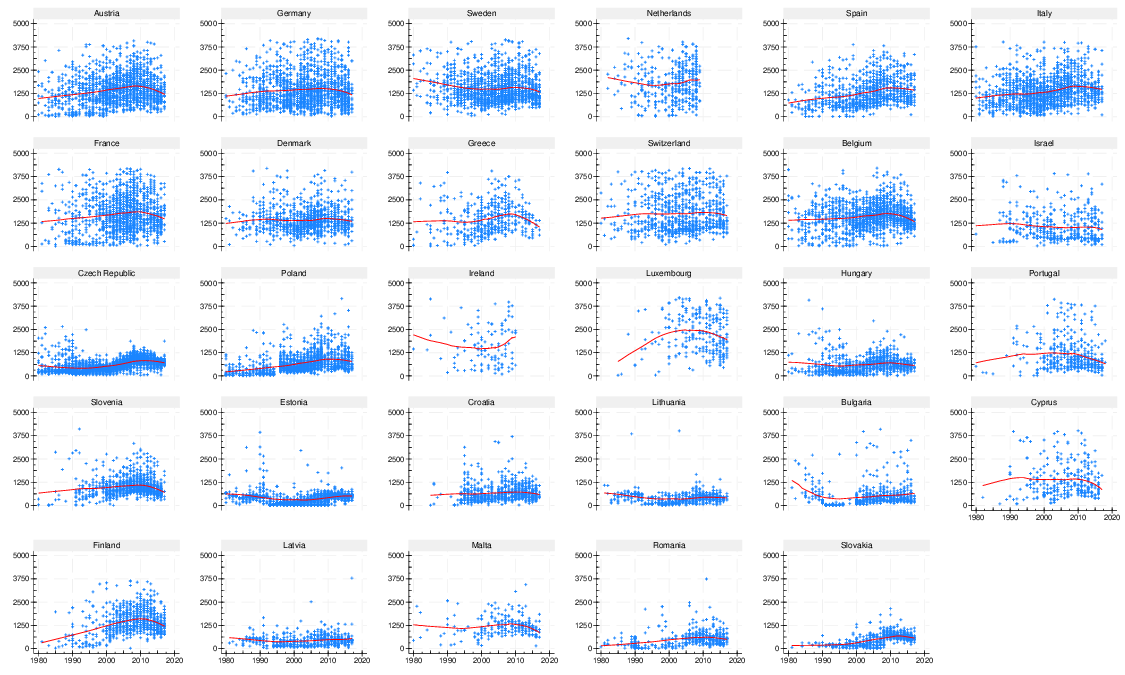}
\caption{Scatterplot of monthly pension benefits and year of retirement}
\label{fig:pbr}
\end{figure}

\begin{figure}[htp]
\centering
\includegraphics[width=1.0\linewidth,height=0.40\textheight]{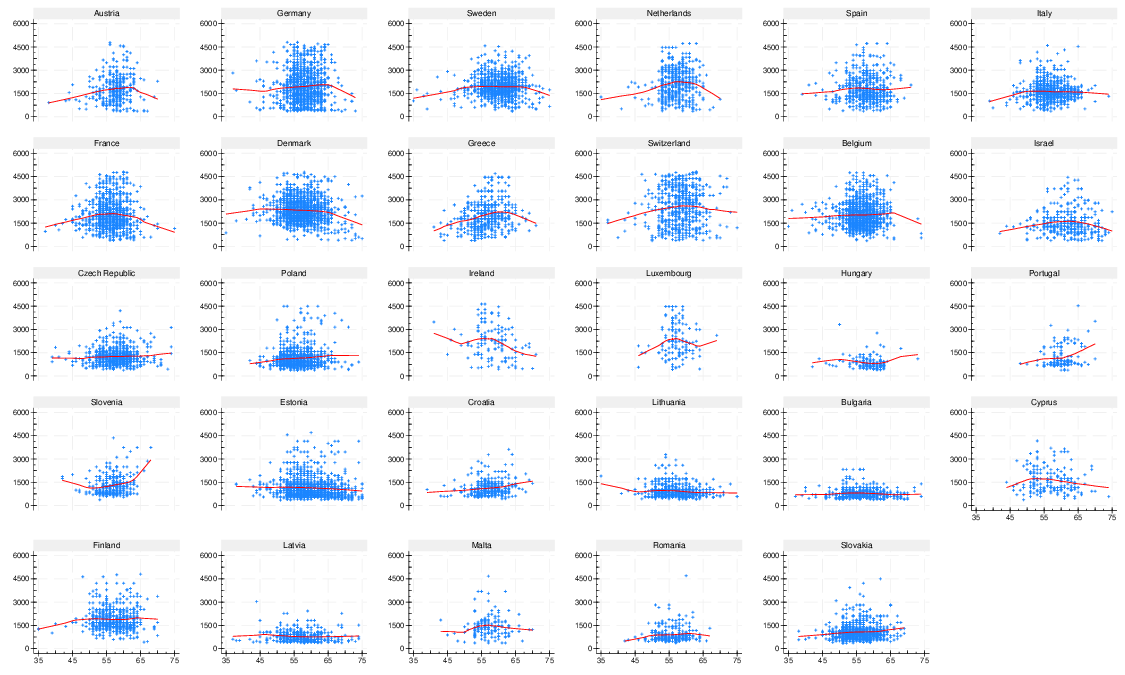}
\caption{Scatterplot between current monthly wages and individual age}
\label{fig:scatter-cmw}
\end{figure}

\begin{figure}[htp]
\centering
\includegraphics[width=1.0\linewidth,height=0.40\textheight]{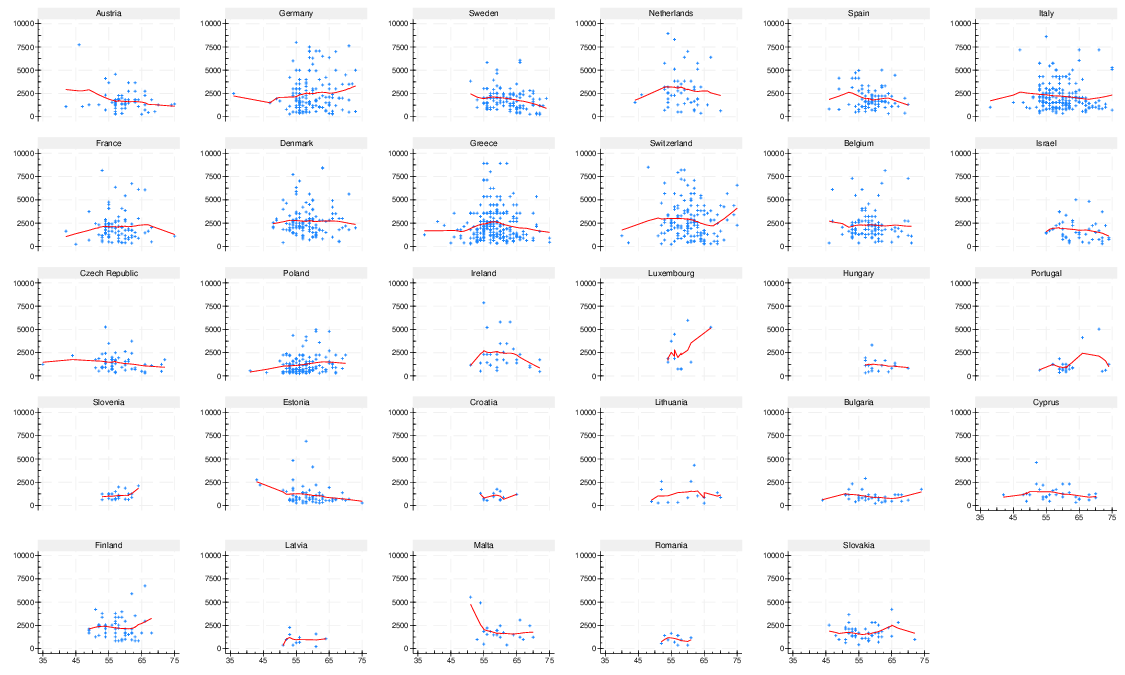}
\caption{Scatterplot between current monthly incomes from self-employment and individual age}
\label{fig:scatter-cmi}
\end{figure}

\begin{figure}[htp]
\centering
\includegraphics[width=1.0\linewidth,height=0.40\textheight]{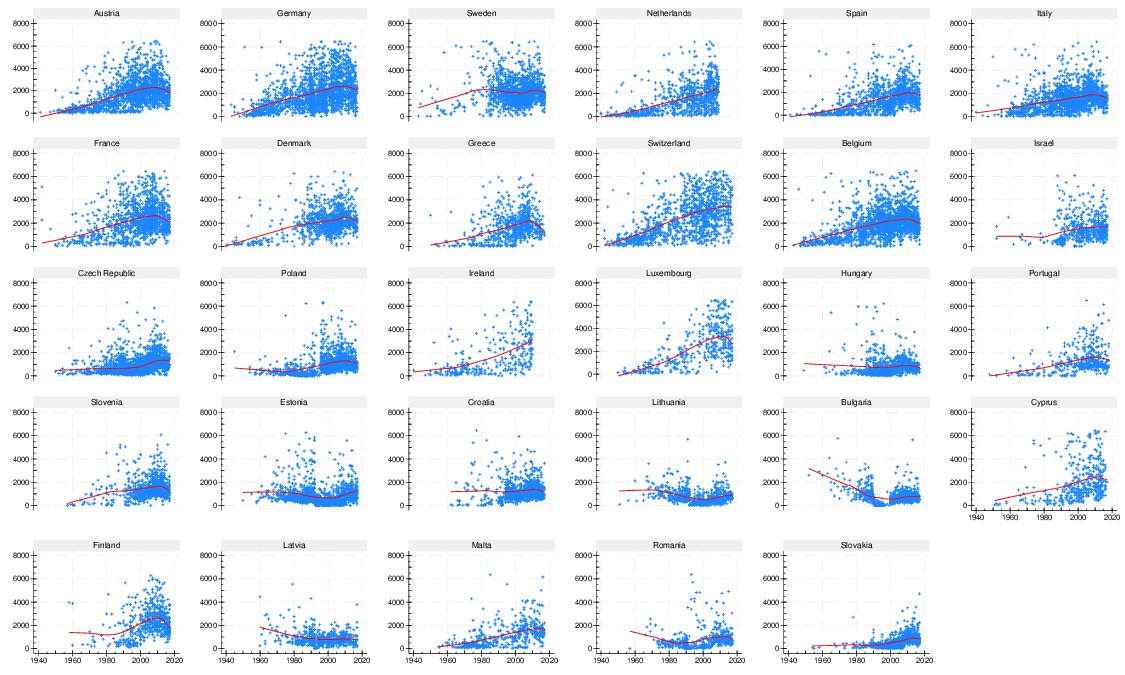}
\caption{Scatterplot of monthly wages at the end of the main job over time}
\label{fig:scatter-mjob-wage}
\end{figure}

\begin{figure}[htp]
\centering
\includegraphics[width=1.0\linewidth,height=0.40\textheight]{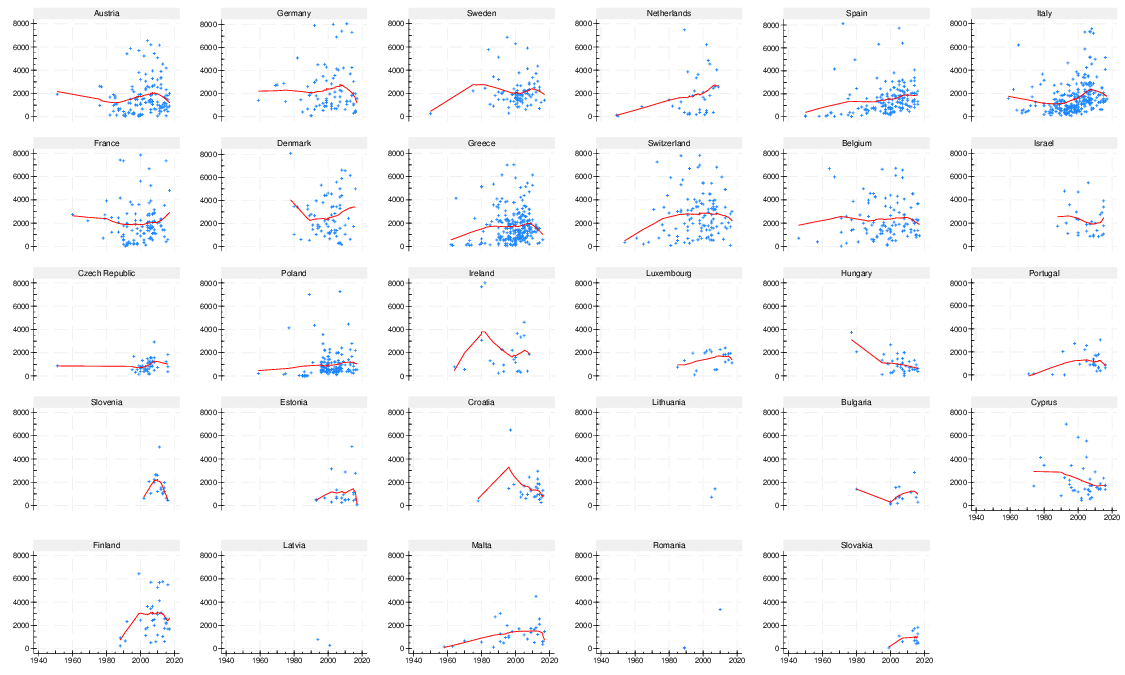}
\caption{Scatterplot of monthly incomes from self-employment at the end of the main job over time}
\label{fig:scatter-mjob-inc}
\end{figure}
\clearpage

\newpage
\section*{Appendix C. Inverse propensity score weighting}

This appendix provides a brief overview of the implementation of the inverse propensity score weighting (IPW) approach.

Tables~\ref{tab:IPW_mb}--\ref{tab:IPW_mjob_income} report the ML estimates for logit models of the response probability for the different monetary outcomes by macro-region.
In addition to the general set of predictors described in Section~\ref{sec:MI_predictors}, each model includes a second-order polynomial in the respondent’s age at the reference period of the outcome:
childbirth for maternity benefits,
the start of job spells for first monthly wages and first monthly incomes from self-employment,
and the end of the last job spell for pension benefits, as well as for monthly wages and incomes at the end of the main job.
This specification of the response model is coherent with the choice of exogenous predictors for the imputation of monetary variables.
The ML estimates of these logit models are used to predict the response probability for each outcome and to compute the propensity score weights, defined as the inverse of the estimated response probabilities.

\begin{figure}[htp]
\centering
\includegraphics[width=1.0\linewidth,height=0.40\textheight]{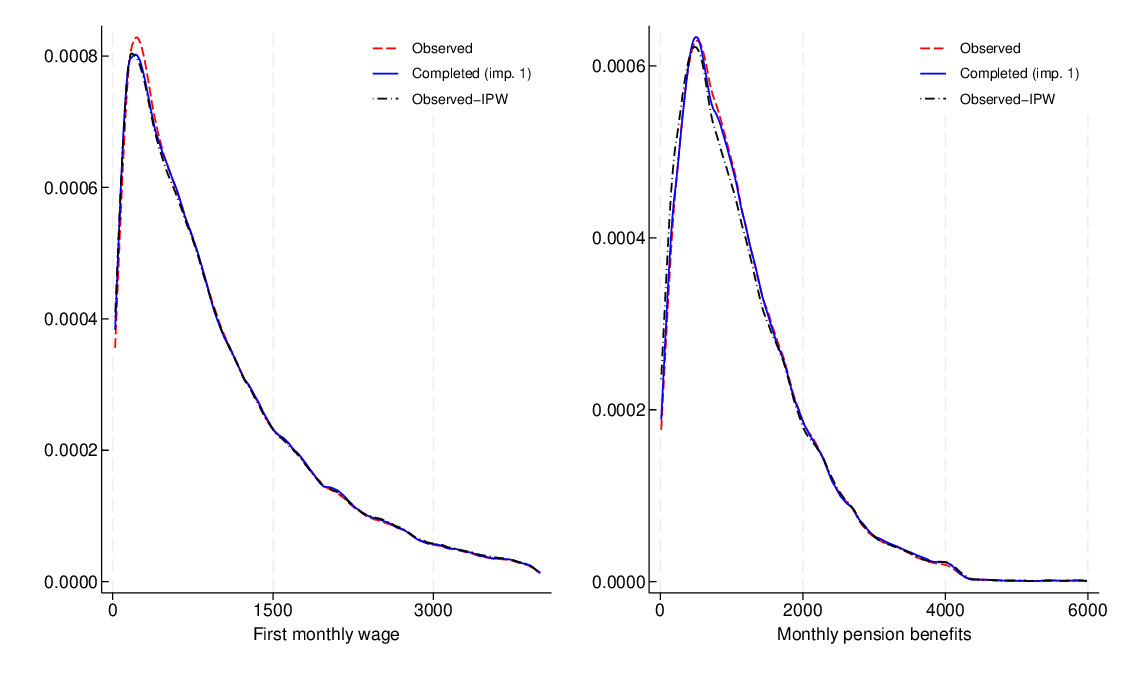}
\caption{Kernel densities of first monthly wages and monthly pension benefits in the observed,
completed, and IPW-weighted samples}
\label{fig:kden2_wage_pbr}
\end{figure}

\begin{figure}[htp]
\centering
\includegraphics[width=1.0\linewidth,height=0.40\textheight]{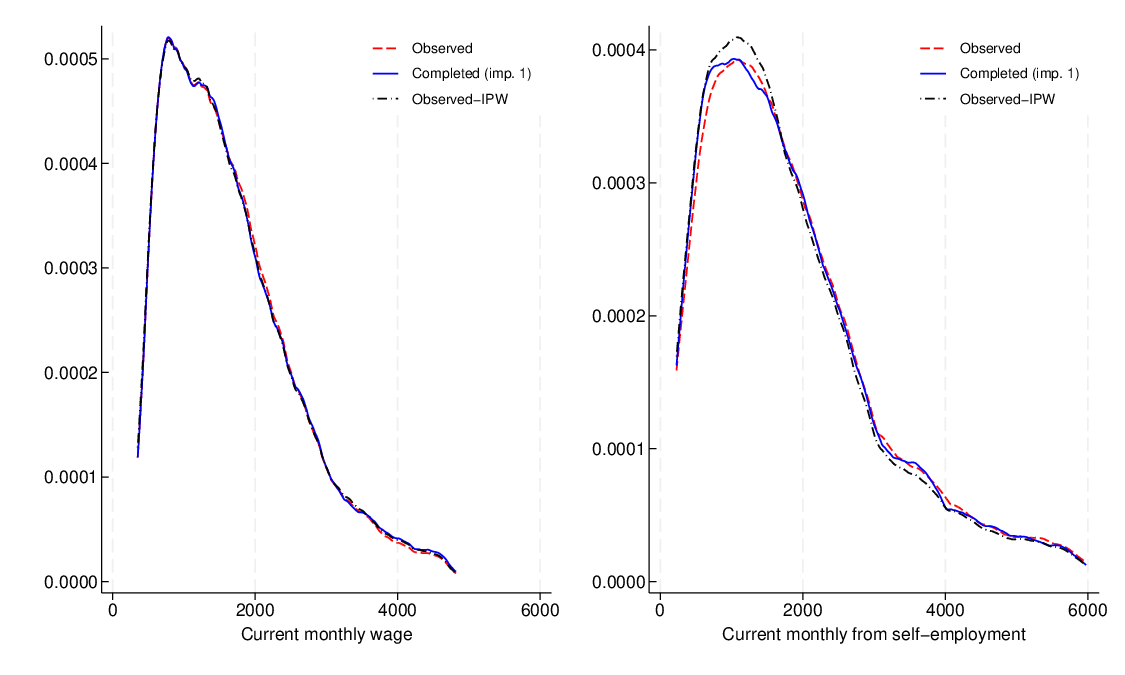}
\caption{Kernel densities of current monthly wages and current monthly incomes from self-employment
in the observed, completed, and IPW-weighted samples}
\label{fig:kden2_cjob_wage_income}
\end{figure}

\begin{figure}[htp]
\centering
\includegraphics[width=1.0\linewidth,height=0.40\textheight]{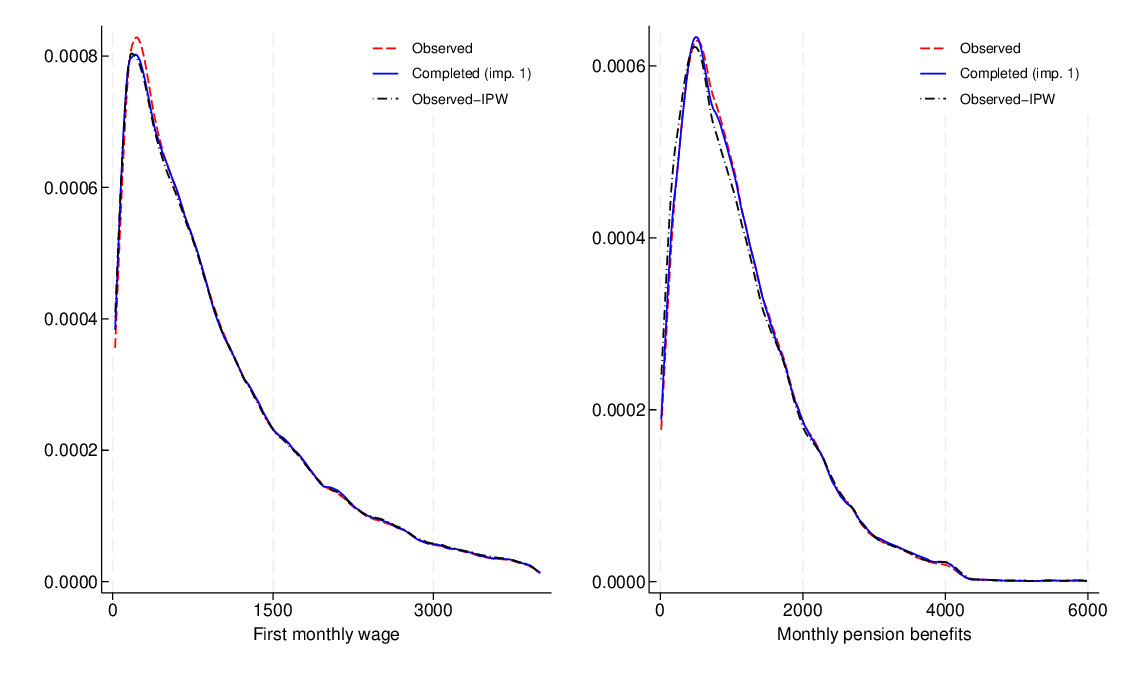}
\caption{Kernel densities of monthly wages and monthly incomes from self-employment at the end of main job
in the observed, completed, and IPW-weighted samples}
\label{fig:kden2_mjob_wage_income}
\end{figure}

Comparisons of kernel density plots in the observed, completed, IPW-weighted samples
are presented in Figures~\ref{fig:kden2_wage_pbr}, \ref{fig:kden2_cjob_wage_income}, and~\ref{fig:kden2_mjob_wage_income}.
Specifically, Figure~\ref{fig:kden2_wage_pbr} focuses on first monthly wages and monthly pension benefits,
Figure~\ref{fig:kden2_cjob_wage_income} focuses on current monthly wages and current monthly income from self-employment,
while
Figure~\ref{fig:kden2_mjob_wage_income} focuses on the monthly wages and monthly income from self-employment at the end of the main job.
For the kernel density plots of monthly maternity benefits and first monthly income from self-employment see
Figure~\ref{fig:kden2_mb_income}.

\begin{table}
\caption{ML estimates of logit models for the response probability of monthly maternity benefits by macro-region}
\begin{small}
\begin{center}
\begin{tabular}{lr@{\,}lr@{\,}lr@{\,}lr@{\,}lr@{\,}lr@{\,}l}
\hline
& \multicolumn{12}{c}{Macro regions} \\
\cline{2-13}
\multicolumn{1}{l}{Variable} &
\multicolumn{2}{c}{Saxons} &
\multicolumn{2}{c}{Nordic} &
\multicolumn{2}{c}{Mixed} &
\multicolumn{2}{c}{Central} &
\multicolumn{2}{c}{Medit.} &
\multicolumn{2}{c}{Balkans} \\
\hline
Age at childbirth             &  -.024&  &  -.099&* &  -.074&  &  -.055&  &  -.096&  &  -.076&  \\
(Age at childbirth)$^2$       &   .000&  &   .002&* &   .001&  &   .001&  &   .002&  &   .002&  \\
Age at interview year         &  -.097&  &   .170&**&  -.090&* &  -.202&**&   .010&  &  -.111&* \\
(Age at interview year)$^2$   &   .001&  &  -.001&**&   .001&* &   .001&**&  -.000&  &   .001&  \\
Couple                        &   .002&  &   .248&**&  -.122&  &  -.138&  &  -.419&  &  -.094&  \\
Household size                &  -.012&  &   .030&  &  -.059&  &  -.019&  &  -.111&  &  -.025&  \\
\hline
\end{tabular}
\end{center}
\end{small}
\label{tab:IPW_mb}
\begin{center}
\vspace{-0.7cm}
\parbox{141mm}{\footnotesize Notes:
Estimates are based on pooled data from multiple spells.
Symbols: * denotes a p-value between 5\% and 1\%, and ** a p-value below 1\%.}
\end{center}
\end{table}

\begin{table}
\caption{ML estimates of logit models for the response probability of first monthly wages by macro-region}
\begin{small}
\begin{center}
\begin{tabular}{lr@{\,}lr@{\,}lr@{\,}lr@{\,}lr@{\,}lr@{\,}l}
\hline
& \multicolumn{12}{c}{Macro regions} \\
\cline{2-13}
\multicolumn{1}{l}{Variable} &
\multicolumn{2}{c}{Saxons} &
\multicolumn{2}{c}{Nordic} &
\multicolumn{2}{c}{Mixed} &
\multicolumn{2}{c}{Central} &
\multicolumn{2}{c}{Medit.} &
\multicolumn{2}{c}{Balkans} \\
\hline
Age at start of job           &  -.021&**&  -.067&**&  -.114&**&  -.088&**&   .018&**&  -.099&**\\
(Age at start of job)$^2$     &   .000&* &   .001&**&   .002&**&   .001&**&  -.000&  &   .002&**\\
Age at interview year         &   .064&**&   .269&**&   .023&* &  -.027&  &   .060&**&   .209&**\\
(Age at interview year)$^2$   &  -.001&**&  -.002&**&  -.000&**&  -.000&  &  -.001&**&  -.002&**\\
Female                        &  -.367&**&  -.057&  &  -.341&**&  -.326&**&  -.210&**&  -.199&**\\
Couple                        &   .142&**&   .101&**&   .034&  &   .206&**&   .080&* &  -.016&  \\
\hline
\end{tabular}
\end{center}
\end{small}
\label{tab:IPW_jsp_wage}
\begin{center}
\vspace{-0.7cm}
\parbox{148mm}{\footnotesize Notes:
Estimates are based on pooled data from multiple spells.
Symbols: * denotes a p-value between 5\% and 1\%, and ** a p-value below 1\%.}
\end{center}
\end{table}

\begin{table}
\caption{ML estimates of logit models for the response probability of first monthly incomes from self-employment by macro-region}
\begin{small}
\begin{center}
\begin{tabular}{lr@{\,}lr@{\,}lr@{\,}lr@{\,}lr@{\,}lr@{\,}l}
\hline
& \multicolumn{12}{c}{Macro regions} \\
\cline{2-13}
\multicolumn{1}{l}{Variable} &
\multicolumn{2}{c}{Saxons} &
\multicolumn{2}{c}{Nordic} &
\multicolumn{2}{c}{Mixed} &
\multicolumn{2}{c}{Central} &
\multicolumn{2}{c}{Medit.} &
\multicolumn{2}{c}{Balkans} \\
\hline
Age at start of job           &   .067&**&  -.044&  &  -.005&  &   .032&  &   .037&* &   .006&  \\
(Age at start of job)$^2$     &  -.001&* &   .001&  &   .001&  &  -.000&  &  -.000&  &   .000&  \\
Age at interview year         &  -.058&  &   .215&* &  -.155&**&   .006&  &  -.055&  &   .198&  \\
(Age at interview year)$^2$   &   .000&  &  -.002&**&   .001&* &  -.000&  &   .000&  &  -.002&  \\
Female                        &  -.367&**&  -.121&  &  -.433&**&  -.594&**&  -.091&  &  -.314&  \\
Couple                        &  -.110&  &  -.126&  &   .022&  &  -.129&  &  -.085&  &  -.080&  \\
\hline
\end{tabular}
\end{center}
\end{small}
\label{tab:IPW_jsp_income}
\begin{center}
\vspace{-0.7cm}
\parbox{143mm}{\footnotesize Notes:
Estimates are based on pooled data from multiple spells.
Symbols: * denotes a p-value between 5\% and 1\%, and ** a p-value below 1\%.}
\end{center}
\end{table}

\begin{table}
\caption{ML estimates of logit models for the response probability of monthly pension benefits by macro-region}
\begin{small}
\begin{center}
\begin{tabular}{lr@{\,}lr@{\,}lr@{\,}lr@{\,}lr@{\,}lr@{\,}l}
\hline
& \multicolumn{12}{c}{Macro regions} \\
\cline{2-13}
\multicolumn{1}{l}{Variable} &
\multicolumn{2}{c}{Saxons} &
\multicolumn{2}{c}{Nordic} &
\multicolumn{2}{c}{Mixed} &
\multicolumn{2}{c}{Central} &
\multicolumn{2}{c}{Medit.} &
\multicolumn{2}{c}{Balkans} \\
\hline
Age at end of job spell       &   .042&  &   .014&**&   .027&**&   .020&**&   .030&**&   .044&**\\
(Age at end of job spell)$^2$ &  -.000&  &   .000&  &   .000&  &   .000&**&   .000&**&   .001&**\\
Age at interview year         &  -.079&  &  -.042&  &  -.189&**&  -.195&**&   .028&  &  -.417&**\\
(Age at interview year)$^2$   &   .000&  &  -.000&  &   .001&* &   .001&  &  -.001&* &   .002&**\\
Female                        &   .039&  &  -.033&  &  -.124&* &   .062&  &   .026&  &  -.195&* \\
Couple                        &   .090&  &  -.057&  &   .029&  &   .073&  &   .087&  &  -.098&  \\
\hline
\end{tabular}
\end{center}
\end{small}
\label{tab:IPW_pbr}
\begin{center}
\vspace{-0.7cm}
\parbox{145mm}{\footnotesize
Symbols: * denotes a p-value between 5\% and 1\%, and ** a p-value below 1\%.}
\end{center}
\end{table}

\begin{table}
\caption{ML estimates of logit models for the response probability of current monthly wages by macro-region}
\begin{small}
\begin{center}
\begin{tabular}{lr@{\,}lr@{\,}lr@{\,}lr@{\,}lr@{\,}lr@{\,}l}
\hline
& \multicolumn{12}{c}{Macro regions} \\
\cline{2-13}
\multicolumn{1}{l}{Variable} &
\multicolumn{2}{c}{Saxons} &
\multicolumn{2}{c}{Nordic} &
\multicolumn{2}{c}{Mixed} &
\multicolumn{2}{c}{Central} &
\multicolumn{2}{c}{Medit.} &
\multicolumn{2}{c}{Balkans} \\
\hline
Age at interview year         &   .147&  &   .274&**&   .269&**&   .202&* &   .114&  &   .206&  \\
(Age at interview year)$^2$   &  -.002&* &  -.002&**&  -.003&**&  -.002&**&  -.001&* &  -.003&  \\
Female                        &   .372&**&  -.010&  &   .147&  &   .414&**&  -.071&  &  -.214&  \\
Couple                        &   .076&  &  -.037&  &   .122&  &  -.163&  &  -.069&  &  -.178&  \\
Household size                &  -.091&  &  -.086&  &  -.019&  &   .010&  &   .057&  &   .008&  \\
High education                &   .032&  &  -.067&  &   .079&  &  -.203&  &  -.140&  &   .286&  \\
\hline
\end{tabular}
\end{center}
\end{small}
\label{tab:IPW_cjob_wage}
\begin{center}
\vspace{-0.7cm}
\parbox{145mm}{\footnotesize
Symbols: * denotes a p-value between 5\% and 1\%, and ** a p-value below 1\%.}
\end{center}
\end{table}

\begin{table}
\caption{ML estimates of logit models for the response probability of current monthly incomes from self-employment by macro-region}
\begin{small}
\begin{center}
\begin{tabular}{lr@{\,}lr@{\,}lr@{\,}lr@{\,}lr@{\,}lr@{\,}l}
\hline
& \multicolumn{12}{c}{Macro regions} \\
\cline{2-13}
\multicolumn{1}{l}{Variable} &
\multicolumn{2}{c}{Saxons} &
\multicolumn{2}{c}{Nordic} &
\multicolumn{2}{c}{Mixed} &
\multicolumn{2}{c}{Central} &
\multicolumn{2}{c}{Medit.} &
\multicolumn{2}{c}{Balkans} \\
\hline
Age at interview year         &  -.065&  &   .401&  &  -.068&  &  -.048&  &  -.119&  &   .205&  \\
(Age at interview year)$^2$   &   .000&  &  -.003&  &   .000&  &   .000&  &   .001&  &  -.002&  \\
Female                        &  -.441&* &  -.154&  &  -.246&  &  -.543&**&  -.382&**&  -.144&  \\
Couple                        &  -.071&  &  -.280&  &   .694&**&  -.372&  &  -.348&  &   .215&  \\
Household size                &  -.099&  &  -.165&* &  -.220&  &   .181&  &   .033&  &   .333&  \\
High education                &  -.058&  &  -.031&  &   .773&**&  -.104&  &   .166&  &   .275&  \\
\hline
\end{tabular}
\end{center}
\end{small}
\label{tab:IPW_cjob_income}
\begin{center}
\vspace{-0.7cm}
\parbox{143mm}{\footnotesize
Symbols: * denotes a p-value between 5\% and 1\%, and ** a p-value below 1\%.}
\end{center}
\end{table}

\begin{table}
\caption{ML estimates of logit models for the response probability of monthly wages at the end of the main job by macro-region}
\begin{small}
\begin{center}
\begin{tabular}{lr@{\,}lr@{\,}lr@{\,}lr@{\,}lr@{\,}lr@{\,}l}
\hline
& \multicolumn{12}{c}{Macro regions} \\
\cline{2-13}
\multicolumn{1}{l}{Variable} &
\multicolumn{2}{c}{Saxons} &
\multicolumn{2}{c}{Nordic} &
\multicolumn{2}{c}{Mixed} &
\multicolumn{2}{c}{Central} &
\multicolumn{2}{c}{Medit.} &
\multicolumn{2}{c}{Balkans} \\
\hline
Age at end of job spell       &   .010&  &  -.009&  &  -.011&  &  -.011&  &   .017&  &  -.123&**\\
(Age at end of job spell)$^2$ &  -.000&  &   .000&* &   .000&**&   .000&**&  -.000&  &   .002&**\\
Age at interview year         &   .050&  &  -.028&  &   .001&  &   .012&  &   .027&  &  -.117&**\\
(Age at interview year)$^2$   &  -.001&* &  -.000&  &  -.000&  &  -.001&* &  -.001&* &   .000&  \\
Female                        &   .059&  &  -.065&  &  -.084&  &  -.050&  &  -.115&* &  -.091&  \\
Couple                        &   .182&* &   .010&  &   .031&  &   .151&* &   .151&* &  -.044&  \\
\hline
\end{tabular}
\end{center}
\end{small}
\label{tab:IPW_mjob_wage}
\begin{center}
\vspace{-0.7cm}
\parbox{145mm}{\footnotesize
Symbols: * denotes a p-value between 5\% and 1\%, and ** a p-value below 1\%.}
\end{center}
\end{table}

\begin{table}
\caption{ML estimates of logit models for the response probability of monthly incomes from self-employment at the end of the main job by macro-region}
\begin{small}
\begin{center}
\begin{tabular}{lr@{\,}lr@{\,}lr@{\,}lr@{\,}lr@{\,}lr@{\,}l}
\hline
& \multicolumn{12}{c}{Macro regions} \\
\cline{2-13}
\multicolumn{1}{l}{Variable} &
\multicolumn{2}{c}{Saxons} &
\multicolumn{2}{c}{Nordic} &
\multicolumn{2}{c}{Mixed} &
\multicolumn{2}{c}{Central} &
\multicolumn{2}{c}{Medit.} &
\multicolumn{2}{c}{Balkans} \\
\hline
Age at end of job spell       &   .007&  &   .015&  &  -.076&  &   .085&  &   .032&  &  -.000&  \\
(Age at end of job spell)$^2$ &   .000&  &  -.000&  &   .001&  &  -.001&  &  -.000&  &   .001&  \\
Age at interview year         &  -.018&  &  -.293&* &  -.169&  &  -.059&  &   .073&  &  -.135&  \\
(Age at interview year)$^2$   &   .000&  &   .002&  &   .001&  &   .000&  &  -.001&  &   .000&  \\
Female                        &  -.245&  &  -.391&* &  -.492&* &  -.220&  &  -.219&  &  -.008&  \\
Couple                        &   .062&  &  -.064&  &   .547&  &   .001&  &   .099&  &   .798&  \\
\hline
\end{tabular}
\end{center}
\end{small}
\label{tab:IPW_mjob_income}
\begin{center}
\vspace{-0.7cm}
\parbox{148mm}{\footnotesize
Symbols: * denotes a p-value between 5\% and 1\%, and ** a p-value below 1\%.}
\end{center}
\end{table}

\end{document}